\documentclass{article}

\pdfoutput=1

\usepackage{jheppub}
\usepackage{epsfig}
\usepackage{setspace}
\usepackage{subfigure}
\usepackage{amssymb,amsmath}
\usepackage{graphicx}
\usepackage{color}
\usepackage{cancel}

\title{Higgs Mass and Muon Anomalous Magnetic Moment in the MSSM with
Gauge-Gravity Hybrid Mediation}

\author[1]{Bin Zhu}
\author[2]{Ran Ding}
\author[3,4,5]{Tianjun Li}

\affiliation[1]{Department of Physics, Yantai University, Yantai 264005, P. R. China}

\affiliation[2]{Center for High-Energy
Physics, Peking University, Beijing, 100871, P. R. China}

\affiliation[3]{Key Laboratory of Theoretical Physics and Kavli Institute for Theoretical Physics China (KITPC),
Institute of Theoretical Physics, Chinese Academy of Sciences, Beijing 100190, P. R. China}

\affiliation[4]{ School of Physical Sciences, University of Chinese Academy of Sciences,
  Beijing 100049, P. R. China}

\affiliation[5]{School of Physical Electronics, University of Electronic Science and Technology of China,
Chengdu 610054, P. R. China}

\date{\today}

\emailAdd{zhubin@mail.nankai.edu.cn}

\abstract{
 In general, we can propose the hybrid supersymmetry breakings and hybrid mediations in the
Supersymmetric Standard Models (SSMs). In this paper,
we study the hybrid mediation for supersymmetry (SUSY) breaking. In particular, we study
how to keep the good properties of gravity mediation, gauge mediation, and anomaly mediation,
while solve their problems simultaneously. As an example, we consider the gauge-gravity mediation, where
all the supersymmetric particles (sparticles) obtain the SUSY breaking soft terms from the traditional gravity mediation
while gauge mediation gives dominant contributions to the soft terms in the colored sector due to the splitted messengers.
Thus, we can realize the electroweak supersymmetry naturally where the sleptons, sneutrinos, and electroweakinos are light within
one TeV while the squarks and gluino are heavy around a few TeVs. And then we can explain 125 GeV Higgs mass, satisfy
the LHC SUSY search bounds, and explain the anomalous magnetic momement of muon, etc.
Moreover, the gluino and squarks are well beyond the current LHC Run II searches.
}

\begin{document}

\maketitle

\section{Introduction}

It is well-known that Supersymmetry (SUSY) provides a natural solution to the
gauge hierarchy problem in the Standard Model (SM). For the supersymmetric SMs (SSMs)
with $R$-parity, gauge coupling unification can be realized, the Lightest
Supersymmetric Particle (LSP) serves as a viable dark matter (DM) candidate,
and electroweak (EW) gauge symmetry can be broken radiatively due to the
large top quark Yukawa coupling, etc. Moreover, gauge coupling unification
strongly suggests Grand Unified Theories (GUTs), and SUSY GUTs can be elegantly
constructed from superstring theory. Therefore, supersymmetry is an important bridge
between the most promising new physics beyond the SM and the high-energy fundamental physics.

The great success at the LHC so far is the discovery of a SM-like Higgs boson $(h)$ with
mass $m_h=125.09\pm 0.24$~GeV~\cite{ATLAS, CMS}. However, to achieve such a SM-like
Higgs boson mass in the Minimal SSM (MSSM), we need either the multi-TeV top squarks with
small mixing or TeV-scale top squarks with large mixing~\cite{Carena:2011aa}. Also,
the LHC SUSY searches give stringent constraints on the viable parameter space of the SSMs.
For example, the latest SUSY search bounds show that the gluino (${\tilde g}$) mass is
heavier than about 1.9 TeV, whereas the light stop (${\tilde t}_1$) mass is heavier than
about 900 GeV~\cite{WAdam-ICHEP}. Thus, there exists the electroweak fine-tuning problem
in the SSM. And then how to construct the natural SSMs, which can realize the
correct Higgs boson mass and satisfy the LHC SUSY search constraints, is a big challenge.

In addition to gauge hierarchy problem in the SM, the anomalous magnetic moment of the muon $a_{\mu}=(g_{\mu}-2)/2$,
remains one of strong hints for physics beyond the SM (BSM) since it is deviated from the SM prediction
 more than $3\sigma$ level. The discrepancy compared to its SM theoretical value is given
as~\cite{Bennett:2006fi, Bennett:2008dy, Davier:2010nc}
\begin{align}
\Delta a_{\mu}=(a_{\mu})_{\text{exp}}-(a_{\mu})_{\text{SM}}
=(28.6\pm 8.0)\times 10^{-10}~.
\end{align}
In the SSMs, the light smuon, muon-sneutrino, Bino, Winos, and Higgsinos would contribute
to $\Delta a_{\mu}$~\cite{Moroi:1995yh,Martin:2001st,Byrne:2002cw,Stockinger:2006zn,Domingo:2008bb}.
The contributions to $\Delta a_{\mu}$ from the neutralino-smuon and chargino-sneutrino loops can approximately be
expressed as
\begin{align}
 \Delta a_{\mu}^{\tilde \chi^0 \tilde \mu} &\simeq \frac{1}{192\pi^2} \frac{m_{\mu}^2}{M_{SUSY}^2} \left({\rm sgn(\mu M_1)}g_1^2-{\rm sgn(\mu M_2)}g_2^2\right)\tan \beta~, \\
 \Delta a_{\mu}^{\tilde \chi^{\pm} \tilde \nu} &\simeq {\rm sgn(\mu M_2)}\frac{1}{32\pi^2} \frac{m_{\mu}^2}{M_{SUSY}^2} g_2^2 \tan \beta~.
 \end{align}
where $M_{SUSY}$ denotes the typical mass scale of the relevant sparticles.
 It is obviously that if all the relevant sparticles have masses around the same scale, the chargino-sneutrino loop
contributions would be dominant. Thus, we have
$\Delta a_{\mu} \sim 10^{-9} \left(\frac{100~{\rm GeV}}{M_{\rm SUSY}}\right)^2 \tan \beta$ for
sgn$(\mu M_2)>0$. From Ref.~\cite{Byrne:2002cw}, we know that the 2$\sigma$ bound on $\Delta a_{\mu}$ can
be achieved for $\tan \beta=10$ if four relevant
sparticles are lighter than $600-700$ GeV.
While for smaller $\tan \beta$ ($\sim$3), the lighter sparticles ($\lesssim 500$ GeV) are needed.
Therefore, to explain the muon anomalous magnetic moment, we need
the light smuon, muon-sneutrino, Bino, Winos, and Higgsinos.

Inspired by the above LHC Higgs~\cite{moriond2013} and SUSY~\cite{LHC-SUSY} searches,
 the experimental results/constraints
on B physics~\cite{Aaij:2012nna, Buchmueller:2009fn},
 Flavour Changing Neutral Current
(FCNC)~\cite{Barberio:2007cr, Asner:2010qj, Amhis:2012bh},
dark matter relic density from WMAP experiment~\cite{Hinshaw:2012aka}, and
direct dark matter search from XENON100 experiment~\cite{Aprile:2012nq},
one of us with his collaborators proposed the Electroweak Supersymmetry (EWSUSY): the squarks and/or gluino
are heavy around a few TeVs while the sleptons, sneutrinos, Bino,
Winos, and/or Higgsinos are light within one TeV~\cite{Cheng:2012np}.
To realize the EWSUSY in the MSSM, we considered the
Generalized Minimal Supergravity (GmSUGRA)~\cite{Li:2010xr, Balazs:2010ha} with
the non-universal gaugino masses, universal/non-universal scalar masses,
and universal/non-universal trilinear soft terms.
For the later and relevant studies,
see Refs.~\cite{Giudice:2012pf, Ibe:2012qu, Grajek:2013ola,
Endo:2013bba,  Mohanty:2013soa,
Bhattacharyya:2013xba, Akula:2013ioa, Iwamoto:2013mya, Choudhury:2013jpa, Cheng:2013hna, Bhattacharyya:2013xma,
Mummidi:2013hba, Li:2014dna, Bhattacharyya:2015vha, Gogoladze:2015jua, Gogoladze:2016grr}.
Especially, the heavy squarks are prefered by the SUSY FCNC and
CP violation problems. And the light electroweak SUSY
sector is very promising on both model building and phenomenological study.

In short, for the SUSY model building and phenomenology, we still need to explore
the natural SSMs which are consistent with all the latest experimental results.
As we know, supersymmetry can be broken via F-term and D-term, and
there are three main mediation mechanisms for the supersymmetry breaking:
gravity mediation, gauge mediation, and anomaly mediation.
Because each mediation mechanism has its merits and drawbacks, we shall propose
the generic hybrid F-term and D-term supersymmetry breaking and hybrid mediations,
where all the merits can be preserved while the problems can be solved.
The examples for the hybrid supersymmetry breakings are the SSMs with pseudo-Dirac gluino~\cite{Ding:2015wma}
and the UV insensitive anomaly mediation~\cite{UVI-AMSB};
and the examples for the hybrid mediations are the deflected anomaly mediation~\cite{D-AMSB},
mirage mediation~\cite{Mirage-Mediation}, deflected mirage mediation~\cite{Deflected-Mirage-Mediation},
and gravity-gauge mediation~\cite{Kang:2011ny}.
In this paper, we study the hybrid mediation for supersymmetry breaking in general,
and set up the solid foundation for our future research program. In particular, we discuss
how to keep the good properties of gravity mediation, gauge mediation, and anomaly mediation,
while solve their problems simultaneously.
As a concrete example, we consider the gravity-gauge mediation, which are comparable.
Once the messenger scale is taken to be around $10^{16}$ GeV, we obtain
$m_{3/2}\sim M_{GMSB}$, and then  gravity mediation can be comparable with gauge mediation.
This proposal is different from the previous gravity-gauge mediation in Ref.~\cite{Kang:2011ny},
where the gauge mediation is dominant. Moreover,
as we know, to obtain the $125$ GeV Higgs boson mass
 in the minimal content of the Gauge Mediated Supersymmmetry Breaking
(GMSB) which employs a pair of ($5$, $\bar 5$) of $SU(5)$ GUT as messengers, the soft masses
of sleptons and electroweakinos are pushed up to a few TeVs and then we cannot explain
the anomalous magnetic momentum of the muon.
In other words, the splitting between the gluino/squarks and the electroweakinos/sleptons/sneutrinos
is not large enough.
This inspired the GMSB proposal with adjoint
 messengers~\cite{Bhattacharyya:2013xma, Mummidi:2013hba, Bhattacharyya:2015vha,Gogoladze:2015jua, Gogoladze:2016grr},
which can naturally generate such large splitting.
However, a critical problem is that the desired large splitting causes the tachyonic slepton problem.
To solve this problem, people have considered
extra one pair of $SU(5)$ ($5$, $\bar 5$) messenger fields~\cite{Bhattacharyya:2013xma},
or additional universal scalar mass and Bino mass from gravity mediation~\cite{Bhattacharyya:2015vha},
or $U(1)_{B-L}$ D-term contributions~\cite{Gogoladze:2015jua, Gogoladze:2016grr}, etc.
In this paper, we assume that the squarks and gluino obtain the large and dominant SUSY breaking soft terms
from gauge mediation, while the sleptons, sneutrinos, Bino,
Winos, and/or Higgsinos obtain the dominant SUSY breaking soft terms from gravity mediation.
Our proposal is different from the above scenarios: (1) The messenger fields need not be $SU(3)$ adjoint field;
(2) we do not have $SU(2)$ adjoint messenger; (3) we have the universal gaugino mass, scalar mass,
and trilinear term from gravity mediation. Interestingly, such kind of hybrid model inherits
the merits of both gravity and gauge mediations. In particular, the $\mu-B_{\mu}$ problem in the GMSB is solved
due to the Giudice-Masiero mechanism for gravity mediation, and the LSP neutralino can still be the DM candidate.

This paper is organized as follows. In section~\ref{sec:hybrid}, we study the
hybrid supersymmetry breakings and hybrid mediations in general. In section~\ref{sec:model} we layout our model contents with taking
into account some soft spectra in GMSB. The phenomenology aspects especially on muon $g-2$ anomaly
and current LHC constraints is investigated in section~\ref{sec:re}.

\section{Hybrid Supersymmetry Breakings and Hybrid Mediations}
\label{sec:hybrid}

In a supersymmetric theorey, the scalar potential has F-terms and D-terms. Thus, there are two kinds of
supersymmetry breakings: F-term and D-term supersymmetry breakings. Usually, we consider the F-term supersymmetry
breaking. As an example, to propose the SSMs with the pseudo-Dirac gluino and Majorana Wino/Bino, we have studied the
hybrid F-term and D-term supersymmetry breaking before~\cite{Ding:2015wma}.

As we know, there are three major supersymmetry breaking mediation mechanisms:

\begin{itemize}

\item {Gravity Mediated Supersymmetry Breaking }

Gravity mediation has been studied systematically and extensively,
especially, the Minimal Supergravity (mSUGRA) and Constrained MSSM (CMSSM) with
 universal supersymmetry breaking soft terms: gaugino mass, scalar mass, and
trilinear term.
Also, the Generalized mSUGRA (GmSUGRA) can have non-universal soft terms.
Usually, the LSP is the lightest neutralino, and can have correct density via
coannihilation, well-tempered neutralino, and resonance annihilation, etc.
Or we can have multi-component dark matter and then the Higgsino can be the LSP.
In the string models, gravity mediation has been considered as the dominant mediation mechanism as well.
In short, gravity mediation has enough degrees of freedom and can be consistent with
all the current experimental results.
However, in principle,
the big challenge is the Flavor Changing Neutral Current (FCNC) problem and CP violation problem
for the general gravity mediation.

\item{Gauge Mediated Supersymmetry Breaking (GMSB) }

Gauge mediation solves the FCNC and CP violation problems since the gauge interaction is flavour blind.
Unlike gravity mediation, the LSP is gravitino, whose mass can be constrained from
cosmology. For the minimal gauge mediation, we have two problems:
(1) The supersymmetry breaking trilinear $A$ term is very small. Thus, to have the correct Higgs boson
mass, we need heavy stop quarks with mass above 5 TeV. This problem can be solved via the Yukawa deflected
gauge mediation with additional Yukawa mediation; (2) $\mu-B_{\mu}$ problem.

\item{Anomaly Mediated Supersymmetry Breaking (AMSB) }

Anomaly mediation can solve the FCNC and CP violation problems as well. Anomaly mediation
requires a mechanism to suppress gravity mediation since it is generated at loop level.
For example, the sequestering mechanism between the observable and hidden sectors.
The gaugino mass is proportional to the corresponding one-loop gauge beta function,
while scalar masses and trilinear $A$-terms depend on the anomalous dimensions of the
corresponding scalar fields. Thus, it predicts that Wino will be the LSP. However,
the big problem is that sleptons are tachyonic. There are two kinds of solutions:
(1) The UV insensitive anomaly mediation with addition D-term contributions to scalar masses~\cite{UVI-AMSB};
(2) The deflected anomaly mediation with extra gauge mediation~\cite{D-AMSB}.
In this scenario, gravitino is heavier than 25 TeV and will decay before the
big-bang nucleosynthesis (BBN), so we can solve the cosmological moduli problem.

\end{itemize}

As we know, the hybrid mediations for supersymmetry breaking have been studied previously, and let us give
two kinds of examples:
(1) The deflected anomaly mediation~\cite{D-AMSB},
mirage mediation~\cite{Mirage-Mediation}, deflected mirage mediation~\cite{Deflected-Mirage-Mediation} can be considered as the
hybrid gauge and anomaly mediations.
(2) The gravity-gauge mediation was proposed previously~\cite{Kang:2011ny}.
But to evade the FCNC and
CP violation problems, the gravity mediation is suppressed compared to the gauge mediation.

As we pointed out in the Introduction, there exists the electroweak fine-tuning problem in the SSMs
and the LHC supersymmetry searches give strong constraints on the SSMs. Therefore, with the
hybrid supersymmetry breakings and hybrid mediations, we may construct the natural SSMs which
can evade the LHC supersymmetry search constraints and solve the problem in the SSMs with single
mediation. Let us give a few examples: (1) In the Next to MSSM (NMSSM) with gauge mediation,
there is a problem for the singlet ($S$) soft mass and its trilinear term $A_{\kappa} \kappa S^3$.
This problem can be solved by considering the hybrid gravity-gauge mediation;
(2) In the anomaly mediation, we can introduce another superfield or modulus whose F-term is non-zero,
and its supersymmetry breaking effect is mediated to the observable sector via gravity interaction.
Thus, the tachyonic slepton problem in anomaly mediation and the moduli problem in gravity mediation
can be solved simultaneously;
(3) In the natural SSMs with small fine-tuning measures, we generically have the non-universal boundary
conditions, and sometimes there exists a little bit hierarchy between the soft mass terms. For example,
we need the following approximate gaugino mass relation~\cite{Antusch:2012gv}
\begin{align}
\left(\frac{M_1}{15M_3}\right)^2 + \left(\frac{M_2}{2.6M_3}\right)^2 = 1 ~,~\,
\end{align}
which can be realized naturally via hybrid mediation.
(4) In the super-natural supersymmetry~\cite{Leggett:2014mza, Leggett:2014hha, Du:2015una, Li:2015dil},
the electroweak fine-tuning measure is automatically at the
order one, and then we can solve the supersymmetry electroweak fine-tuning problem naturally. To generalize
it to the effective super-natural supersymmetry~\cite{Ding:2015epa}, we need additional supersymmetry breaking soft terms,
which are subdominant and will not affect the electroweak fine-tuning measure.
Thus, we can combine the no-scale supergravity or M-theory inspired supergravity with
the other mediations.

\section{Model}
\label{sec:model}
The most general SUSY breaking soft terms in the MSSM are~\cite{Chung:2003fi},
\begin{align}
\mathcal{L}_{\text{soft}}=&-\frac{1}{2}(M_3\tilde g\tilde g + M_2\tilde W\tilde W + M_1\tilde B\tilde B + {\rm h.c.})\nonumber\\
&-(\tilde q T_u \tilde u H_u-\tilde q T_d \tilde d H_d-\tilde l T_e \tilde e H_d + {\rm h.c.})\nonumber\\
&- m^2_{H_u}H_u^*H_u-m^2_{H_d}H_d^*H_d-(B_{\mu}H_u H_d + {\rm h.c.})\nonumber\\
&-\tilde f^+ m_{f}^2 \tilde f\;.
\end{align}
In the GMSB, SUSY breaking~\cite{Giudice:1998bp} is parameterized by a spurion field $X$ with
\begin{align}
\left<X\right>=M+\theta^2 F\;.
\end{align}
In our model, clearly different from the minimal GMSB, $X$ directly couples to colored messenger multiplet $\Phi_{G}$
which takes adjoint representation under $SU(3)$ gauge group. The corresponding superpotential can be written as
\begin{align}
W_{\text{mess}}=\lambda X \Phi_{G}\Phi_{G}\;.
\label{eqn:type1}
\end{align}
This gives a supersymmetric mass $M$ as well as a non-supersymmetric mass term $F\phi\tilde\phi+{\rm h.c.}$ for its scalar component,
which leads to a splitted squared-mass $M^2\pm F$ for the complex scalar messengers. Such mass splitting in messenger sector
is then communicated to the soft terms in the MSSM through radiative corrections with messenger loops.
The gaugino masses are generated at one loop order with
\begin{align}
M_i=\frac{g_i^2}{16\pi^2} n_G \Lambda a_{i}\;,
\label{eq:softgauge}
\end{align}
where the $n_G$ is the Dykin index of $\Phi_{G}$. Indices $i=1,2,3$ denote  bino, wino and gluino. For $\Phi_G$ messenger, the corresponding Dynkin indices are $a_{1,2}=0$, $a_3=3$. As a consequence, bino and wino do not receive any contribution from messenger loops while gluino obtain large masses from non-zero $a_3$. Meanwhile, scalar masses comes from the two-loop order with
\begin{align}
&m_{\tilde l}^2=m_{\tilde e}^2=m_{H_u}^2=m_{H_d}^2=0\;,\nonumber\\
&m_{\tilde u}^2=m_{\tilde d}^2=\frac{g_3^4}{32 \pi ^4} n_G \Lambda^2  f\left(\frac{\Lambda}{M_{\rm mess}}\right)\;,\nonumber\\
&m_{\tilde q}^2=\frac{3g_3^4}{16 \pi ^4} n_G \Lambda^2 f\left(\frac{\Lambda}{M_{\rm mess}}\right)\;.
\label{eq:softtype1}
\end{align}
In above equations, $M_{\rm mess}$ is the messenger scale, $\Lambda\equiv F/M_{\rm mess}$, and $f$ as well as the following $g$
in Eq.~(\ref{eq:softtype2}) are the loop-functions which can be found in Ref.~\cite{Giudice:1998bp}. In the absence of additional messenger interactions, the soft bilinear term $B_{\mu}$ and trilinear terms $A_{u,d,e}$ are zero at the messenger scale and are generated by running of renormalization group equations (RGEs). From Eq.~(\ref{eq:softtype1}), one found that the soft masses of slepton and color sectors are obtained with distinct contributions at the messenger threshold. This feature naturally guarantees large mass splitting which is favored by muon $a_{\mu}$ anomaly. Nevertheless, if slepton masses do not get additional contribution, it will suffer from dangerous tachyonic problem. For this reason, we incorporate the gravity mediation to avoid tachyonic sleptons. As we know, gravity is usually ignored when considering gauge mediation. While the situation changes if the messenger scale is around $10^{15}$ GeV where the gravity mediation is comparable with gauge mediation. Thus, in such case, gauge mediation is naturally associated with additional gravity mediation. In supergravity mediation, the flavor structure of the soft terms is strongly constrained by flavor changing neutral current processes (FCNCs) which strongly suggests a flavor-universal soft terms,
\begin{align}
\tilde m_{ij}^2=m_0^2 \delta_{ij},~T_{u,d,e}=A_0 Y_{u,d,e},~M_{1,2,3}=m_{1/2}\;.
\label{eqn:grm}
\end{align}
For simplicity, we denote it as type-I model. The basic property of such type mode is that only color sector obtained soft masses at messenger scale. Thus, higgs mass will be sensitive to the general relation between $m_0$ and $\Lambda$. In addition, relatively large $m_0$ is favored in order to stabilize the slepton masses. As another example, we further consider the messenger sector is $\Phi_D$ and
$\Phi_D^c$ whose quantum numbers
under $SU(3)_C\times SU(2)_L\times U(1)_Y$ are $(\mathbf{3}, \mathbf{1}, \mathbf{-1/3})$ and $(\mathbf{\overline{3}}, \mathbf{1}, \mathbf{1/3})$,
respectively. The corresponding superpotential is
\begin{align}
W_{\text{mess}}=\lambda X \Phi_D\Phi_{D}^c\;.
\label{eqn:type2}
\end{align}
Here, we denote it as type-II model. The corresponding non-zero soft terms at  messenger scale are given by
\begin{align}
&M_1=\frac{g_1^2}{40\pi^2} n_d \Lambda g\left(\frac{\Lambda}{M_{\rm mess}}\right)\;,\nonumber\\
&M_3=\frac{g_3^2}{40\pi^2} n_d \Lambda  g\left(\frac{\Lambda}{M_{\rm mess}}\right)\;,\nonumber\\
&m_{\tilde q}^2=\frac{1}{128 \pi ^4}\left(\frac{1}{150} g_1^4 +\frac{4}{3} g_3^4
   \right) n_d \Lambda ^2 f\left(\frac{\Lambda}{M_{\rm mess}}\right),\nonumber\\
&m_{\tilde u}^2=\frac{1}{128 \pi ^4}\left(\frac{8}{75} g_1^4 +\frac{4}{3} g_3^4\right) n_d \Lambda^2 f\left(\frac{\Lambda}{M_{\rm mess}}\right)\;,\nonumber\\
&m_{\tilde d}^2=\frac{1}{128 \pi ^4}\left(\frac{2}{75}g_1^4 +\frac{4}{3}g_3^4
   \right) n_d \Lambda ^2 f\left(\frac{\Lambda}{M_{\rm mess}}\right),\nonumber\\
&m_{\tilde l}^2=m_{H_u}^2=m_{H_d}^2=\frac{3 g_1^4}{128 \pi ^4}  n_d \Lambda ^2 f\left(\frac{\Lambda}{M_{\rm mess}}\right)\;,\nonumber\\
&m_{\tilde e}^2=\frac{g_1^4}{3200 \pi ^4} n_d \Lambda^2 f\left(\frac{\Lambda}{M_{\rm mess}}\right)\;.
\label{eq:softtype2}
\end{align}
One can see that compared with type-I model, this model has extra contributions from $U(1)_Y$ sector. This scenario leads to bino, slepton and higgs sectors also receive  non-zero soft mass at messenger scale. As we will discuss in the next section, which notably improves the slepton tachyonic problem and changes the particle spectra. based on above discussion, the particle spectra in our model are governed by following parameters
\begin{align}
\{m_0,\;m_{1/2},\;A_{0},\;{\rm sign}(\mu),\;\tan\beta,\;\Lambda,\;M_{\rm mess},\;n_{G/d}\}\;.
\label{eq:parameters}
\end{align}
Among them, $n_{G/d}$ is the Dynkin index of the messengers, and thus is not a variable. Also, ${\rm sign}(\mu)$ is obviously not a variable as well. Therefore, our models only have two more parameters $\Lambda$ and $M_{mess}$ compare to the mSUGRA/CMSSM. In other words, we do not have a larger number of parameters. Then an interesting question is what is the minimal version of these scenarios that can work. To our
knowledge, the minimal scenario is that we consider $(\Phi_D, ~\Phi_D^c)$ as messengers and no-scale supergravity. In the no-scale supergravity, $m_0=A_0=B_{\mu}=0$. And then the parameters are
\begin{align}
\{M_{1/2},\;\tan\beta,\;\Lambda,\;M_{mess}\}\;.
\end{align}
There are only four parameters, just like the mSUGRA/CMSSM.

Next, we would like to emphasize that the generic vector-like
particles, which do not form complete $SU(5)$ or $SO(10)$ representations in GUTs,
can be obtained from the orbifold constructions~\cite{kawa, GAFF, LHYN},
intersecting D-brane model building on Type II orientifolds~\cite{Blumenhagen:2005mu, Cvetic:2002pj, Chen:2006ip},
M-theory on $S^1/Z_2$ with Calabi-Yau compactifications~\cite{Braun:2005ux, Bouchard:2005ag}, and
F-theory with $U(1)$ fluxes~\cite{Donagi:2008ca, Beasley:2008dc, Beasley:2008kw, Donagi:2008kj}.
In particular, the generic vector-like particles from orbifold GUTs
and F-theory GUTs have been studied previously in details in Refs.~\cite{Li:2010hi, Jiang:2009za, Li:2009cy}.
To be concrete, the vector-like particles $(\Phi_D, ~\Phi_D^c)$ in Type-II model can be realized in all the above model buildings,
while the messenger $\Phi_G$ in Type-I model can only be obtained in the intersecting D-brane model building
on Type II orientifolds.

\begin{figure}[!htbp]
\begin{center}
\includegraphics[width=0.45\textwidth]{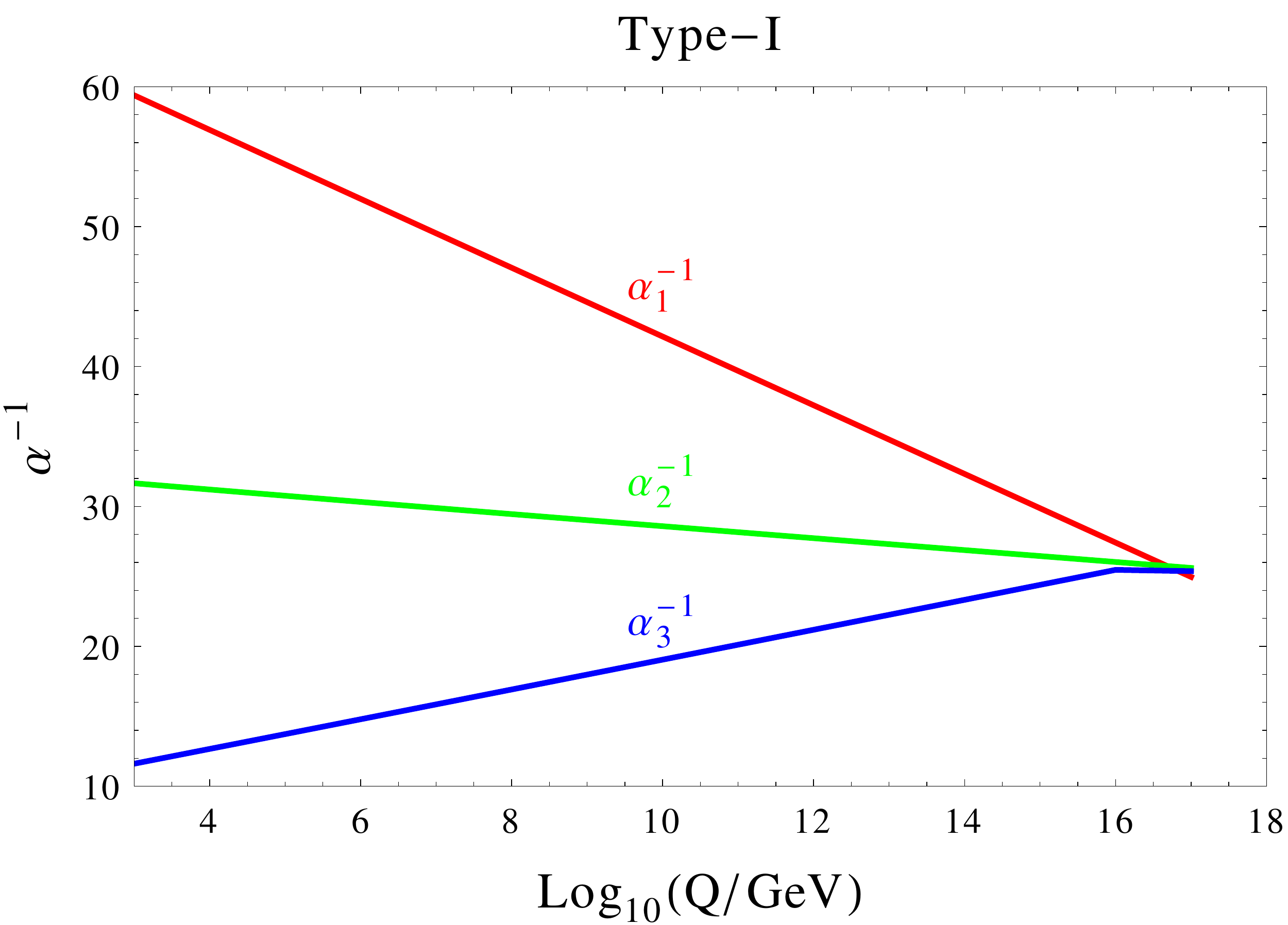}
\includegraphics[width=0.45\textwidth]{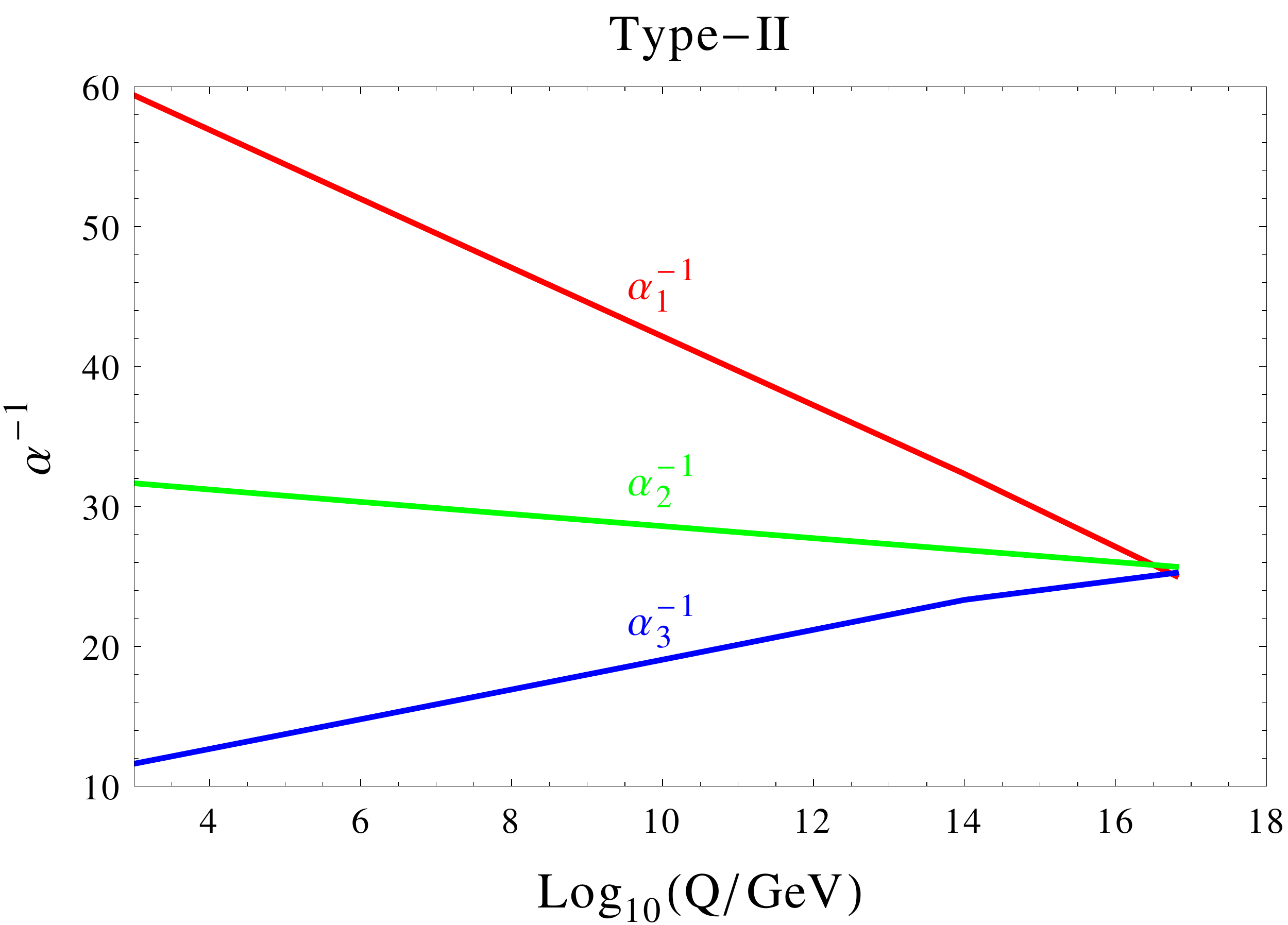}
\end{center}
\caption{Gauge coupling unification in Type-I (left) and Type-II (right) models.
For both two models, GUT can be realized at scale $M_U\simeq 10^{16-17}$ GeV.}
\label{fig:GUT}
\end{figure}

Finally, we briefly comment on realization of the gauge coupling unification (GUT) in both models. As is well known,
the modification of messenger representation plays a crucial role on GUT. Unlike the mGMSB where the messengers form
complete $SU(5)$ representations, $g_3$ coupling has big contributions from
the messenger threshold in Type-I model, which in general violate the conventional GUT condition unless the messenger scale is close to
the GUT scale. While for Type-II model, the situation is improved due to the relatively small contributions to $b_{1,3}$. As a consequence,
the messengers can be at slightly lower scale. The gauge coupling unificatons in Type-I and Type-II models are given in Fig.~\ref{fig:GUT}.

\section{Numerical results}
\label{sec:re}

In this section, we present the numerical studies for both type-I and II models with focus on sparticle spectra and muon anomalous magnetic moment.
In our numerical calculations, the universal soft terms in Eq.~(\ref{eqn:grm}) are firstly generated at GUT scale in terms of gravity mediation for both type-I and type-II models. The behaviors of soft terms then controlled by RGE effects running from GUT scale to messenger scale. Below messenger scale, the messenger fields are integrated out to generate threshold soft terms which are given in Eqs.~(\ref{eq:softgauge}) and (\ref{eq:softtype1})
for Type-I model and in Eq.~(\ref{eq:softtype2}) for Type-II model. For this purpose, we implemented boundary conditions in
Eqs.~(\ref{eqn:grm}),~(\ref{eq:softgauge}),~(\ref{eq:softtype1}), and~(\ref{eq:softtype2}) in the Mathematica package {\tt SARAH}~\cite{Staub:2008uz,Staub:2009bi,Staub:2010jh,Staub:2012pb,Staub:2013tta}. Then {\tt SARAH} is used to create a {\tt SPheno}~\cite{Porod:2003um,Porod:2011nf} version for the MSSM to calculate particle spectrum. The tasks of parameter scans
are implemented by package~{\tt SSP}~\cite{Staub:2011dp}.

\begin{figure}[!htbp]
\begin{center}
\includegraphics[width=0.45\textwidth]{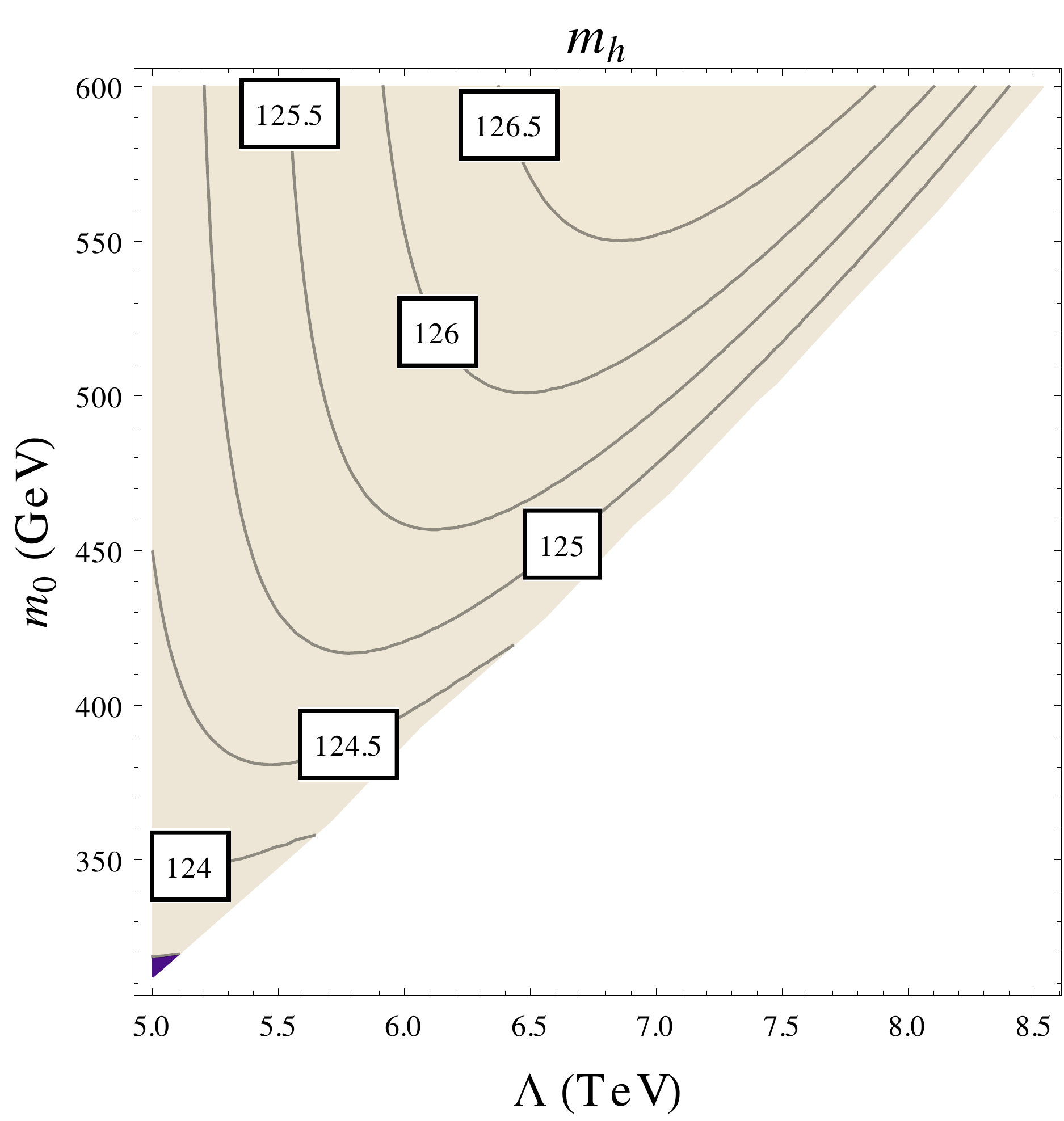}
\includegraphics[width=0.45\textwidth]{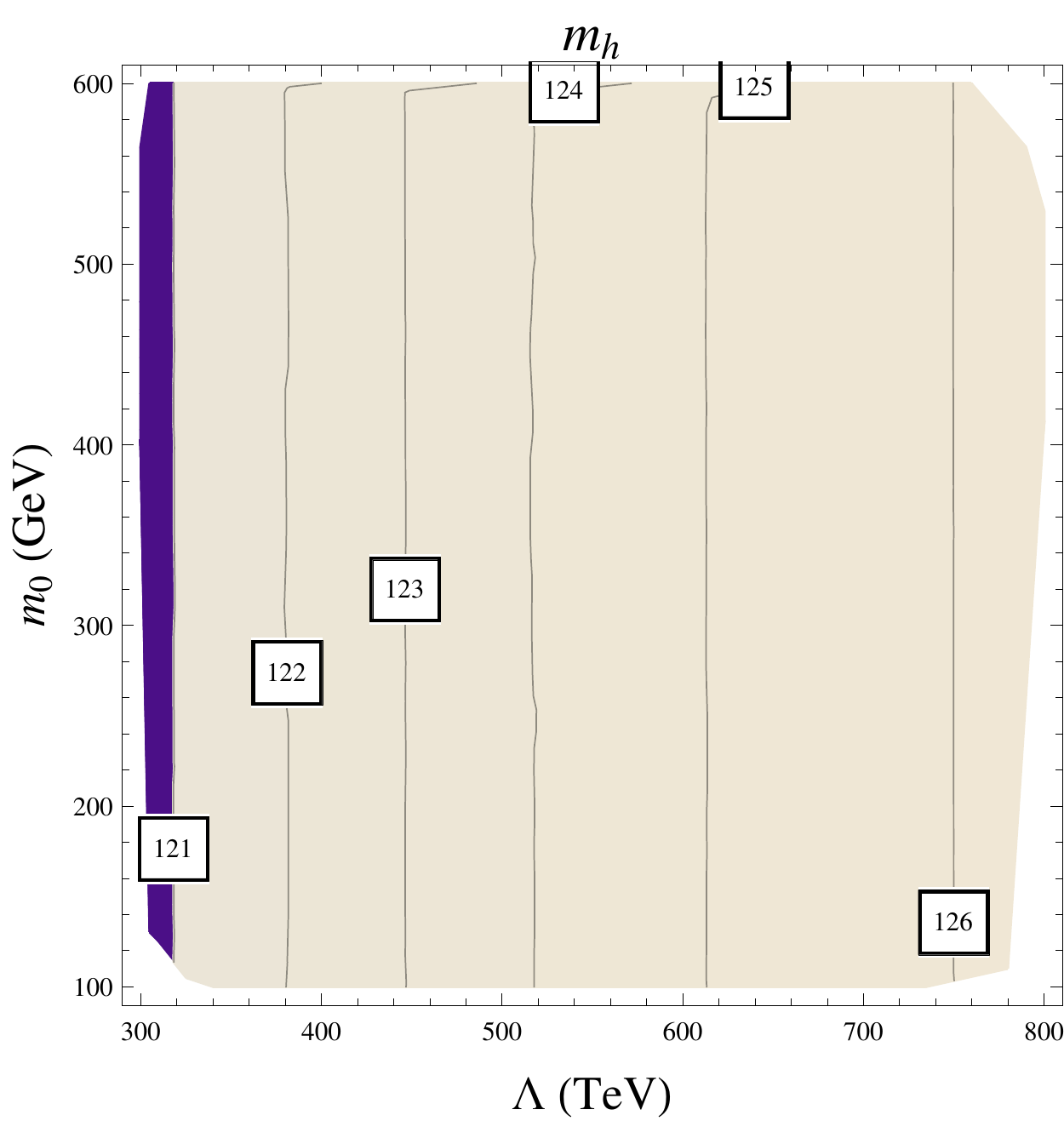}
\end{center}
\caption{Contour distributions of higgs mass on the $\Lambda-m_0$ plane, left (right) panel for type-I (type-II) model.}
\label{fig:higgs}
\end{figure}

\begin{figure}[!htbp]
\begin{center}
\includegraphics[width=0.45\textwidth]{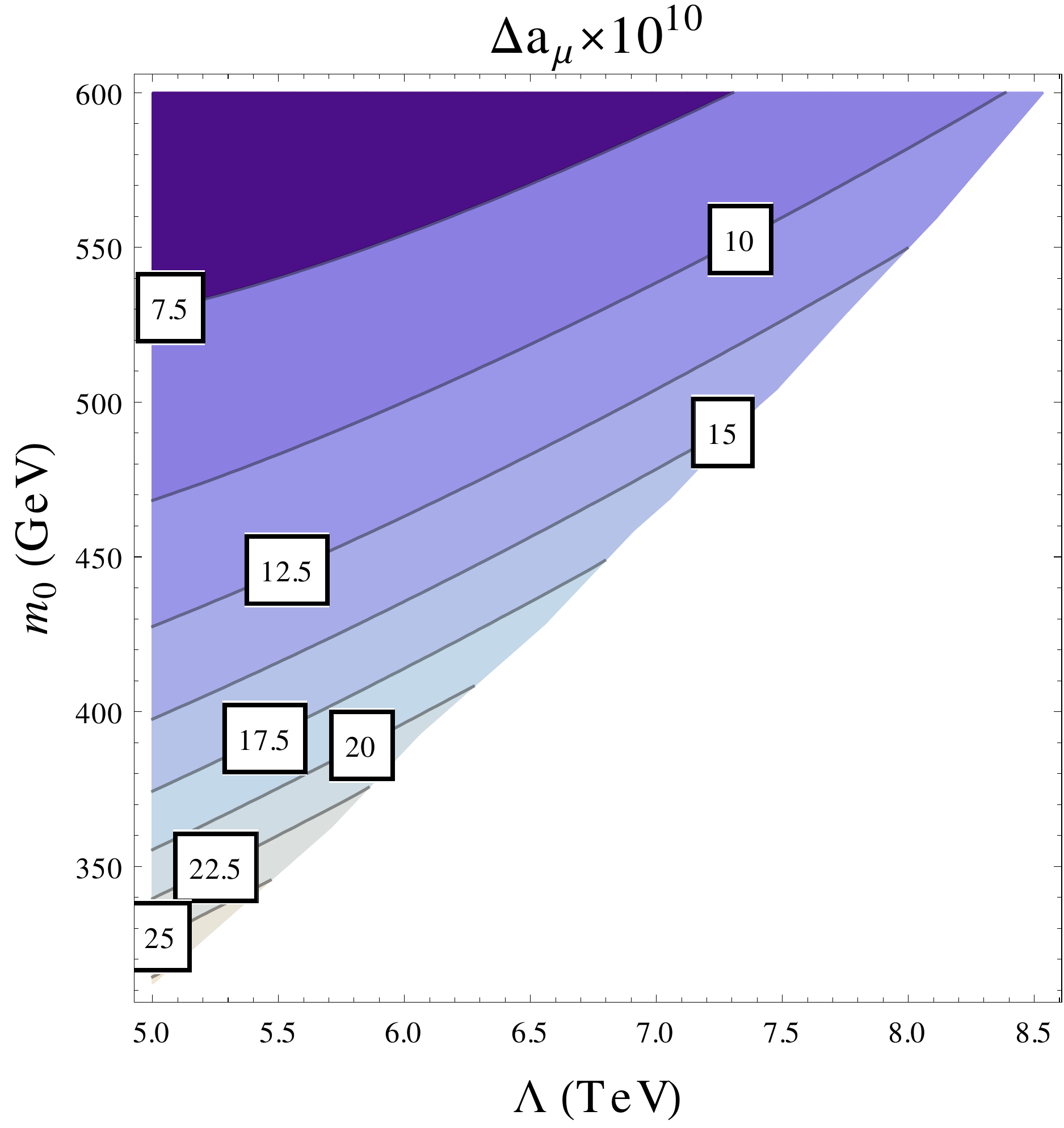}
\includegraphics[width=0.45\textwidth]{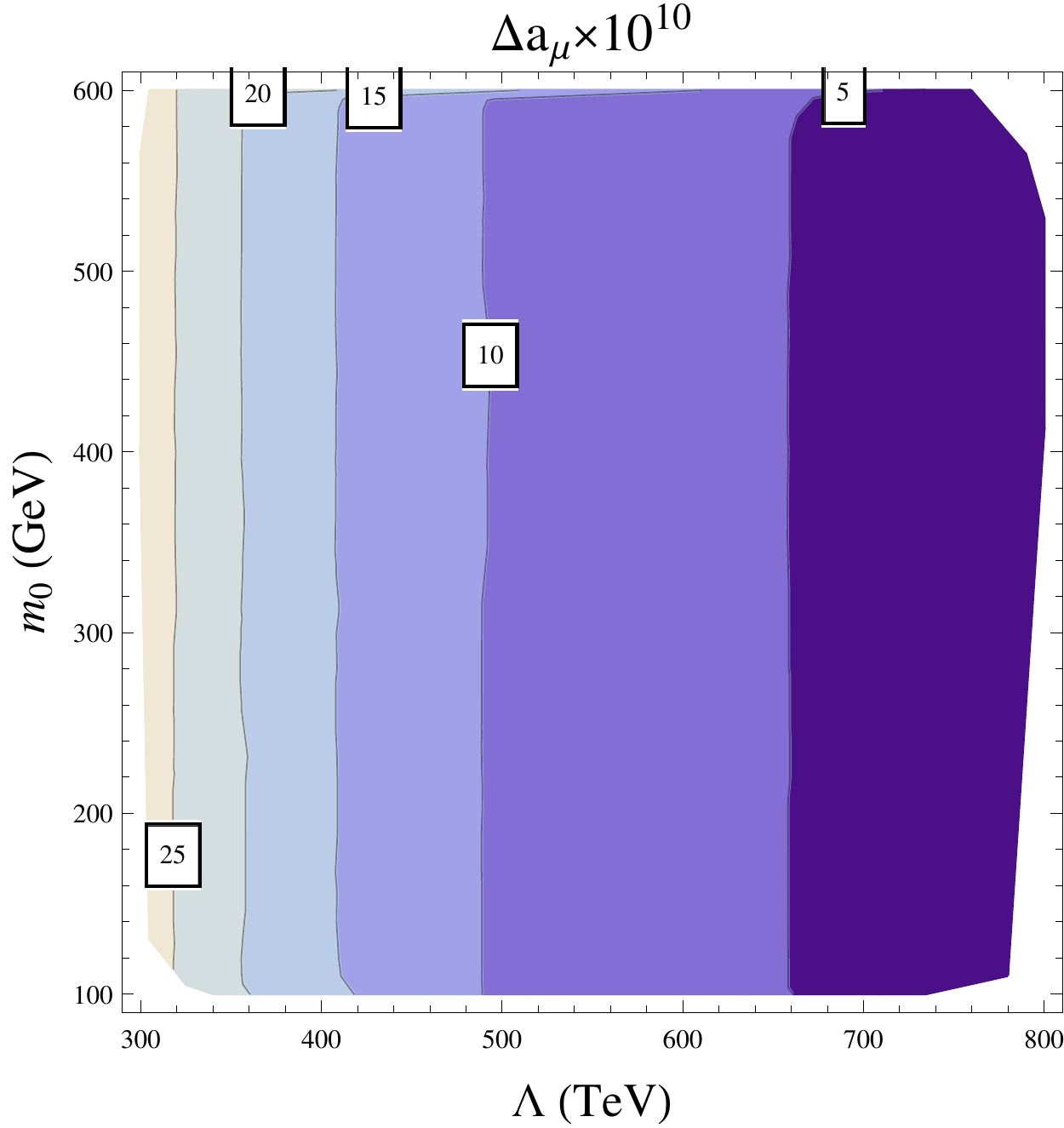}
\end{center}
\caption{Contour distributions of $\Delta a_\mu\times 10^{10}$ on the $\Lambda-m_0$ plane, left (right) panel for type-I (type-II) model.}
\label{fig:g2}
\end{figure}

\begin{figure}[!htbp]
\begin{center}
\includegraphics[width=0.45\textwidth]{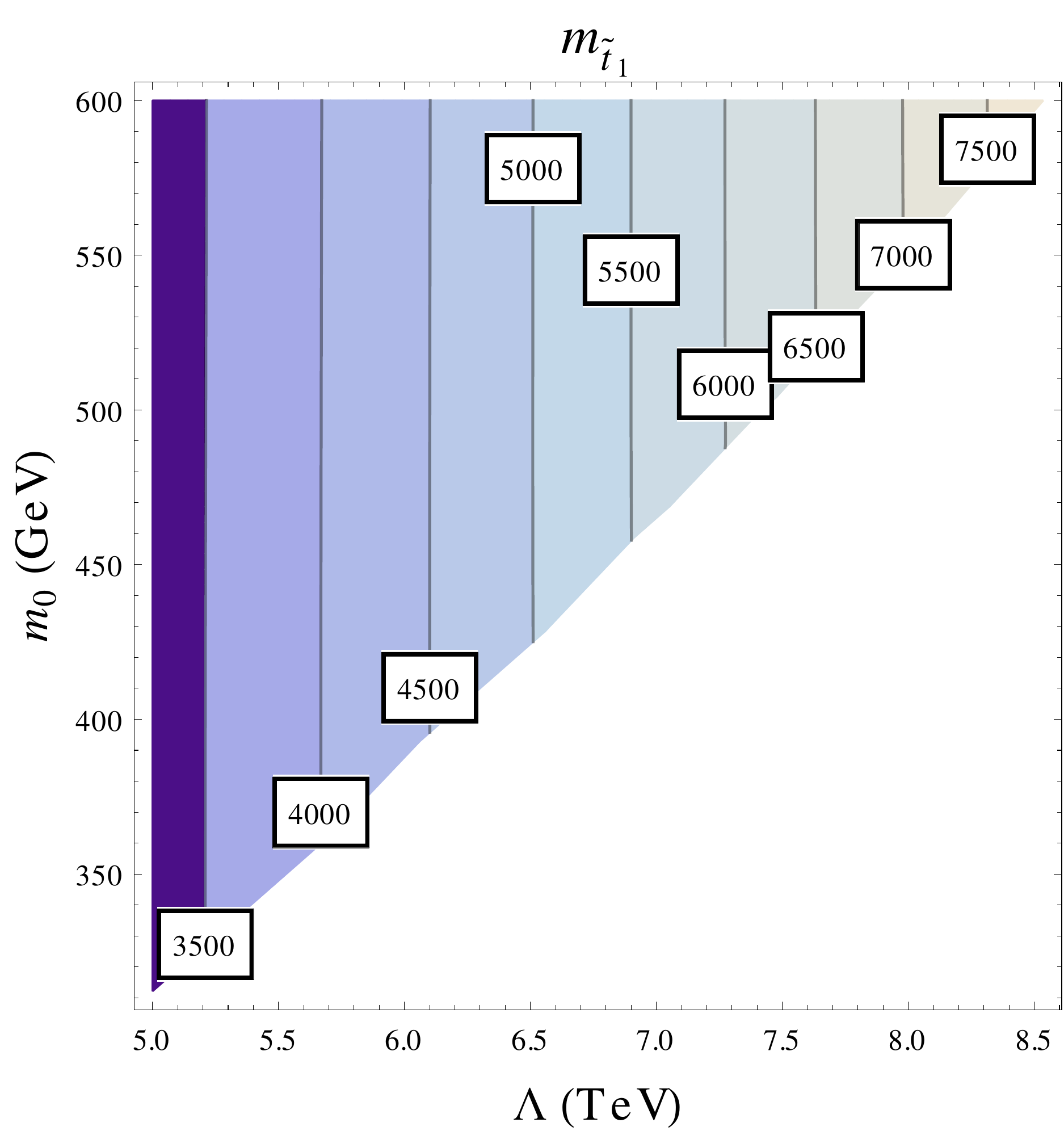}
\includegraphics[width=0.45\textwidth]{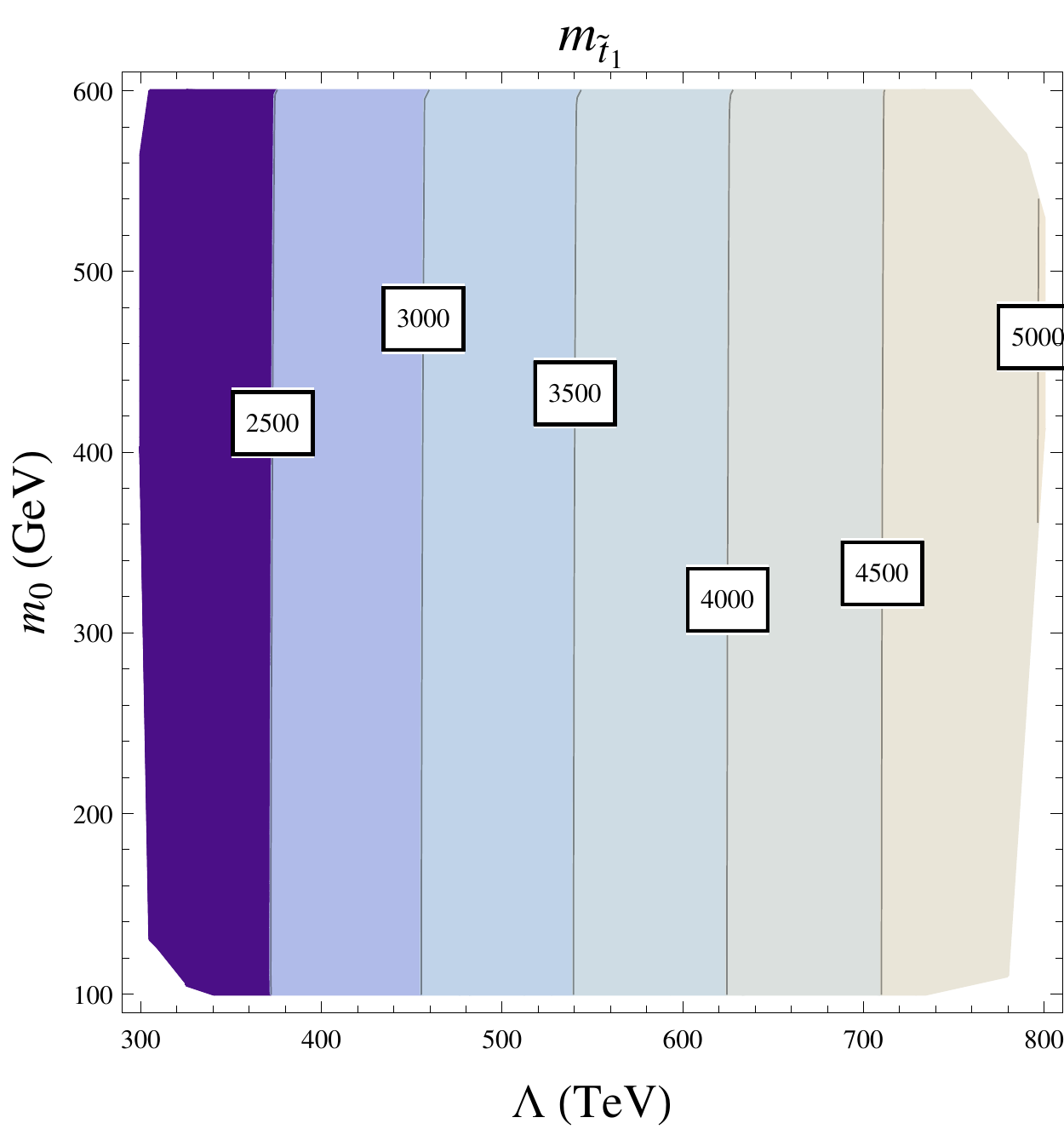}
\label{fig:stop1}
\end{center}
\caption{Contour distributions of light stop mass on the $\Lambda-m_0$ plane, left (right) panel for type-I (type-II) model.}
\end{figure}

\begin{figure}[!htbp]
\begin{center}
\includegraphics[width=0.45\textwidth]{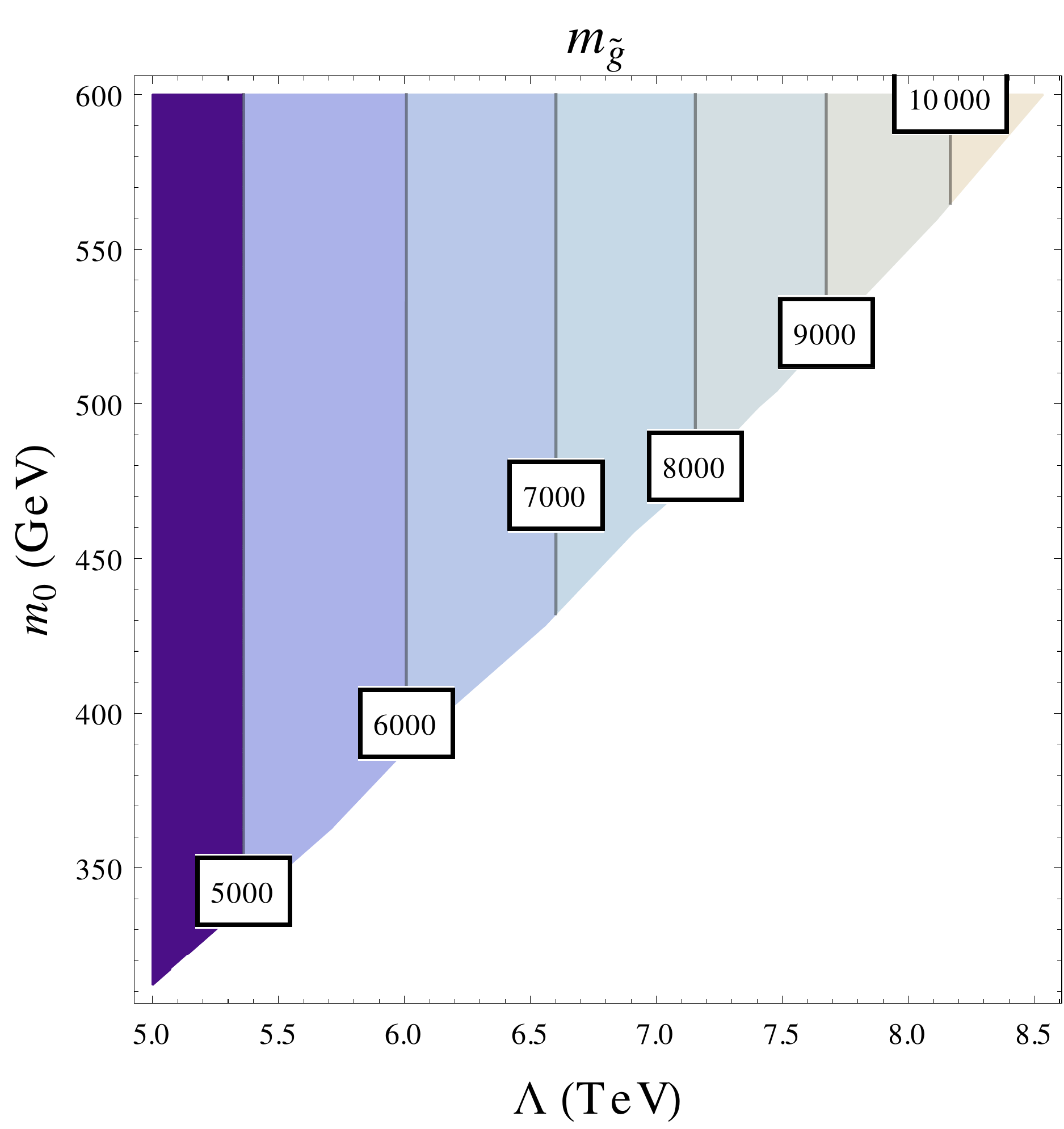}
\includegraphics[width=0.45\textwidth]{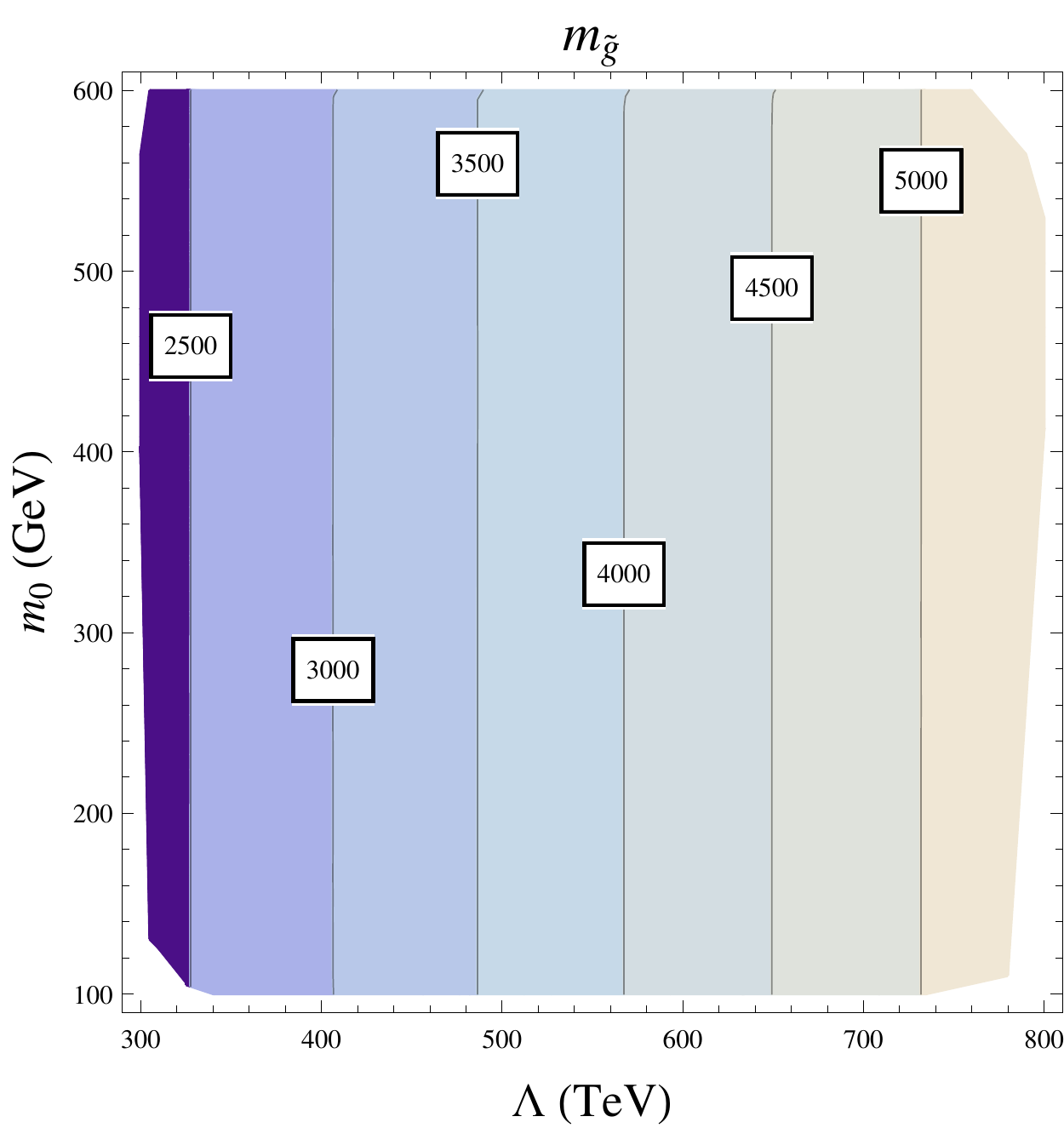}
\label{fig:gluino1}
\end{center}
\caption{Contour distributions of gluino mass on the $\Lambda-m_0$ plane, left (right) panel for type-I (type-II) model.}
\end{figure}

Based on the discussion in section~\ref{sec:model}, the two most important parameters in our model are $m_0$ and $\Lambda$. We thus investigate their effects in first. The contour distributions of $m_h$, $\Delta a_{\mu}$, $m_{{\tilde t}_1}$ and $m_{\tilde g}$ in the $\Lambda-m_0$ are presented in Figs.~\ref{fig:higgs}-\ref{fig:gluino1}, for all of figures, left (right) panel corresponding to type-I (type-II) model. Here, the other parameters are fixed as $M_{mess}=10^{16}$~GeV,
$m_{1/2}=300$~GeV, $A_0=0$, $\tan\beta=20$, and ${\rm sign}(\mu)=1$.

One finds that two type models have distinct behaviors. For type-I model, there exists a boundary which can be fitted by $m_0 \geq 0.08 \Lambda - 95$ GeV. Below this line, parameter space is filled with tachyonic sleptons. On the other hand, type-II model does not suffer from tachyonic problem in most of parameter regions.

Such behaviors can be understood by a simple analysis of RGE effects. Although we use complete two-loop RGEs in numerical calculation. The one-loop RGEs are sufficient to illustrate here. The relevant RGEs are collected in Ref.~~\cite{Martin:1993zk}. Those for slepton masses are given as:
\begin{eqnarray}
\beta_{m^2_{\tilde l}}&=&(m^2_{\tilde l}+2m^2_{H_d})y^\dag_e y_e + 2 y^\dag_e m^2_{\tilde e} y_e + y^\dag_e y_e m^2_{\tilde l} + 2T^\dag_e T_e\nonumber\\ &-&6g^2_2|M_2|^2-\frac{6}{5}g^2_1|M_1|^2-\frac{3}{5}g^2_1\Sigma\;,\\
\beta_{m^2_{\tilde e}}&=&(2m^2_{\tilde e}+4m^2_{H_d})y_e y^\dag_e + 4y_e m^2_{\tilde l} y^\dag_e + 2y_e y^\dag_e m^2_{\tilde e} + 4 T_e T^\dag_e\nonumber\\ &-&\frac{24}{5}g^2_1|M_1|^2+\frac{6}{5}g^2_1\Sigma\;.
\end{eqnarray}
In above equation, quantity $\Sigma$ is defined as
\begin{eqnarray}
\Sigma=m^2_{H_u}-m^2_{H_d}+ {\rm Tr}[m^2_{\tilde q}+m^2_{\tilde l}-m^2_{\tilde u}+m^2_{\tilde d}+m^2_{\tilde e}]\;.
\label{eq:scalar}
\end{eqnarray}
There are four type contributions on the right-hand sides, {\it i.e.}, the terms respectively proportional to Yukawa coupling martix $y_e$, soft trilinear couplings $T_e$, gaugino masses $M_i$ and quantity $\Sigma$. Among them, the first term is only important for third generation, the second and third terms are suppressed in type-I model due to $M_{1,2}=A_0=0$ at messenger scale. In this case, evolution of slepton masses are dominated by $\Sigma$ terms. (for third generation sleptons are dominated by both $\Sigma$ and $y_e$ terms). Once $\Sigma$ has large positive value, the right-handed sleptons especially for third generation are most easy to become tachyons as RGEs running down from the input scale to the low-energy scale. Since quantity $\Sigma$ depends on all the scalar and higgs squared-mass parameters, we present their relevant RGEs as follows
\begin{align}
\beta_{m^2_{\tilde q}}&=&(m^2_{\tilde q}+2m^2_{H_u})y^\dag_u y_u + (m^2_{\tilde q}+2m^2_{H_d})y^\dag_d y_d +[y^\dag_u y_u + y^\dag_d y_d]m^2_{\tilde q} + 2y^\dag_u m^2_{\tilde u}y_u \nonumber\\
&+& 2y^\dag_d m^2_{\tilde d}y_d + 2T^\dag_u T_u + 2T^\dag_d T_d
-\frac{32}{3}g^2_3|M_3|^2-6g^2_2|M_2|^2-\frac{2}{15}g^2_1|M_1|^2 + \frac{1}{5}g^2_1\Sigma\;,\\
\beta_{m^2_{\tilde u}}&=&(2m^2_{\tilde u}+4m^2_{H_u})y_u y^\dag_u + 4y_u m^2_{\tilde q} y^\dag_u + 2y_u y^\dag_u m^2_{\tilde u} + 4 T_u T^\dag_u\nonumber\\ &-&\frac{32}{3}g^2_3|M_3|^2 -\frac{32}{15}g^2_1|M_1|^2 -\frac{4}{5}g^2_1\Sigma\;,\\
\beta_{m^2_{\tilde d}}&=&(2m^2_{\tilde d}+4m^2_{H_d})y_d y^\dag_d + 4y_d m^2_{\tilde q} y^\dag_d + 2y_d y^\dag_d m^2_{\tilde d} + 4 T_d T^\dag_d\nonumber\\ &-&\frac{32}{3}g^2_3|M_3|^2 -\frac{8}{15}g^2_1|M_1|^2 +\frac{2}{5}g^2_1\Sigma\;,\\
\beta_{m^2_{H_u}}&=&6{\rm Tr}\left[(m^2_{H_u}+m^2_{\tilde q})y^\dag_u y_u + y^\dag_u m^2_{\tilde u} y_u + T^\dag_u T_u\right]\nonumber\\
&-& 6g^2_2|M_2|^2 - \frac{6}{5}g^2_1|M_1|^2 + \frac{3}{5}g^2_1\Sigma\;,\\
\beta_{m^2_{H_d}}&=&{\rm Tr}\left[6(m^2_{H_d}+m^2_{\tilde q})y^\dag_d y_d + 6y^\dag_d m^2_{\tilde d} y_d + 2(m^2_{H_d}+m^2_{\tilde l})y^\dag_e y_e
+ 2y^\dag_e m^2_{\tilde e} y_e +6T^\dag_d T_d + 2T^\dag_e T_e\right]\nonumber\\
&-& 6g^2_2|M_2|^2 - \frac{6}{5}g^2_1|M_1|^2 - \frac{3}{5}g^2_1\Sigma\;.
\end{align}
One finds that $m^2_{H_u}$ and $m^2_{H_d}$ are generally decrease as RGEs running down, which is also the necessary condition to trigger electroweak symmetry breaking (EWSB). On the other hand, squark masses receive significant contribution from term proportional to $|M_3|^2$, thus increasing with RGE evolution from the input scale to low-energy scale. Moreover, in type-I model, only gluino and squarks obtain soft masses at messenger scale which are
respectively proportional to $\Lambda$ and $\Lambda^2$. Therefore, contributions of squark masses are much larger than other terms in Eq.~\ref{eq:scalar} ($m^2_{H_u}-m^2_{H_d}$ and $m^2_{\tilde l}+m^2_{\tilde e}$) and dominate the RGE running of slepton masses. As a consequence, whether sleptons become tachyonic are determined by the competition between input $m_0$ at GUT scale and $\Lambda$ at messenger scale. Which obviously imposes a lower bound on $m_0$ for a given $\Lambda$ and indicates such boundary roughly satisfies a linear relation. On the other hand, for type-II model, the bino, slepton and higgs sectors obtain non-zero soft mass at messenger scale (see Eq.~(\ref{eq:softtype2})). Which then relaxes the slepton tachyonic problem through two ways: Firstly, it enhance the initial slepton masses at messenger scale. Secondly, it improves RGE running behavior through term proportional to $|M_1|^2$.

There are some other important features can be learnt from these figures. We summarize them at below:

Firstly, one expects higgs mass should be simply growth with an increases of $\Lambda$ since the higgs mass is mainly lifted by stop mass at 1-loop. It is indeed true for type-II model as is shown in right of Fig.~\ref{fig:higgs}. While the parameter dependence on higgs mass for type-I model is a little bit more complicate. At the low $\Lambda$ region, higgs mass follows the expected behavior. However, with the increasing of $\Lambda$, allowed parameter space is forced to shift to large $m_0$ region in order to obtain the correct higgs mass and avoiding tachyonic sleptons at the same time. As a consequence, a turning-point is appeared for each contour of higg mass. In addition, due to different input soft masses at messenger scale, type-II model is asked for much larger $\Lambda$ compared with type-I model to reach acceptable higgs mass. Therefore, only large $\Lambda$ regions in the figure can satisfy such requirement.

Secondly, muon $g-2$ anomaly requires that slepton and electroweakino masses fall into $\mathcal{O}(100)$~GeV,
$\Delta a_\mu$ is more dependent on $m_0$ than $\Lambda$ for Type-I model. Moreover, the slope of $\Delta a_\mu$ contour reflects RGE effects which are discussed previously, {\it i.e.}, required $m_0$ increases with increasing of $\Lambda$. Especially, $\Delta a_\mu$ is able to approach its observed central value at small $m_0$ and $\Lambda$ regions.
On the other hand, it is difficult to realize for Type-II model. The reason is that
 $\Delta a_\mu$ favors low $\Lambda$ region, while higgs mass requires large $\Lambda$ region.
Therefore, Type-I model is more favored from the point view of explanation of the muon anomalous magnetic moment.

Thirdly, the spectra of colored sparticles fall into the multi-TeV mass range in our model, which certainly exceeds current and upcoming LHC SUSY limits~\cite{Aad:2015pfx,Golling:2016gvc}. In addition, their parameter dependence are simply for both two type models, {\it i.e.},
monotonically increasing with $\Lambda$ as one expects.

\begin{figure}[!htbp]
\begin{center}
\includegraphics[width=0.45\textwidth]{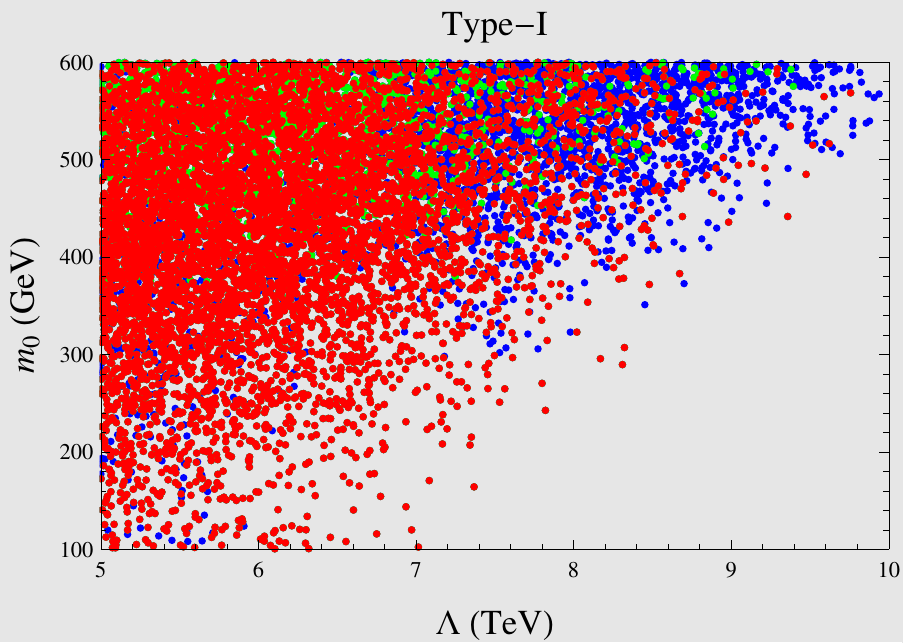}
\includegraphics[width=0.45\textwidth]{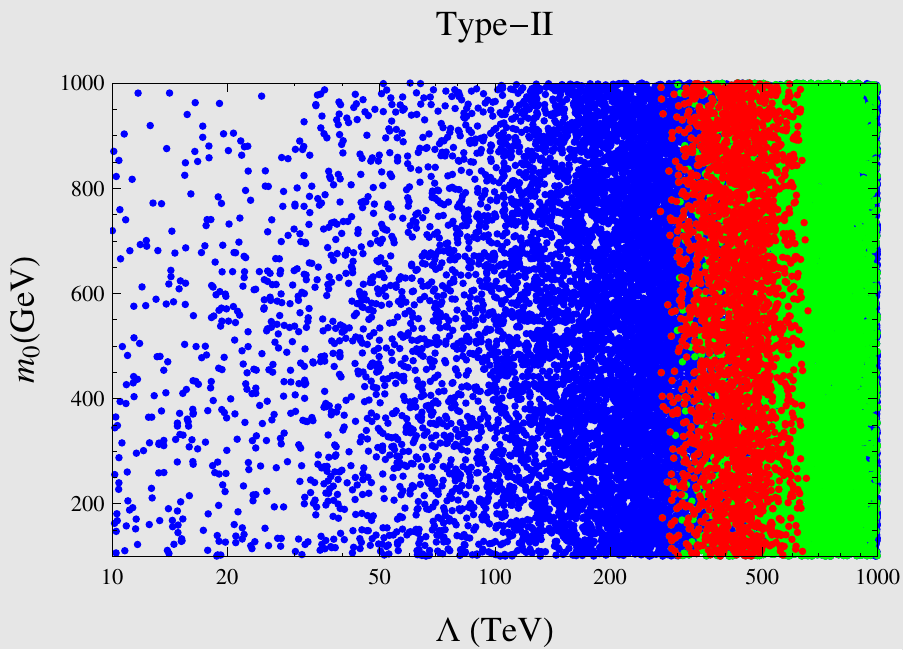}
\end{center}
\caption{Distributions of samples in the $\Lambda-m_0$ plane, left (right) panel for type-I (type-II) model.}
\label{fig:sample1}
\end{figure}

\begin{figure}[!htbp]
\begin{center}
\includegraphics[width=0.45\textwidth]{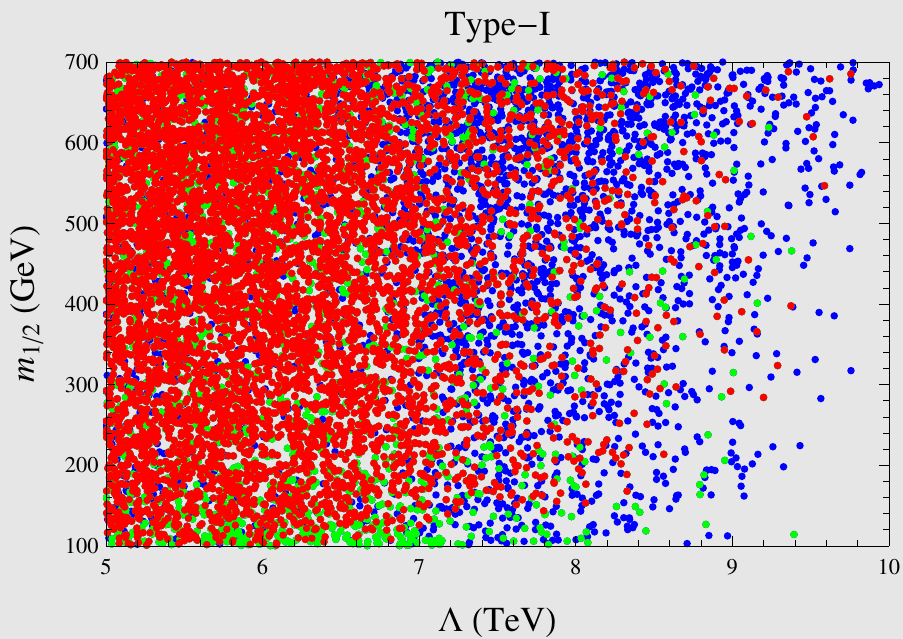}
\includegraphics[width=0.45\textwidth]{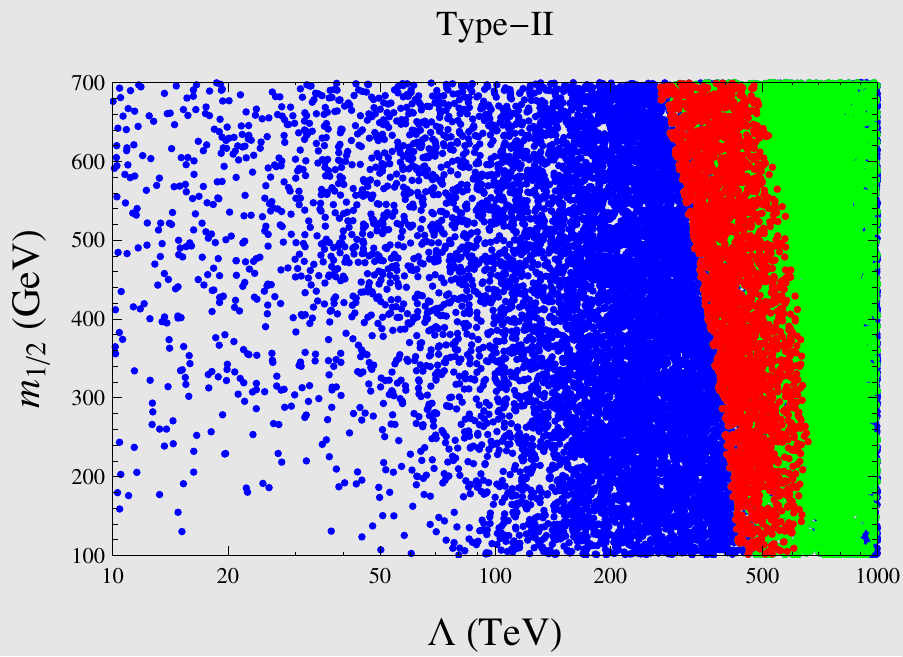}
\end{center}
\caption{Distributions of samples in the $\Lambda-m_{1/2}$ plane, left (right) panel for type-I (type-II) model.}
\label{fig:sample2}
\end{figure}

\begin{figure}[!htbp]
\begin{center}
\includegraphics[width=0.45\textwidth]{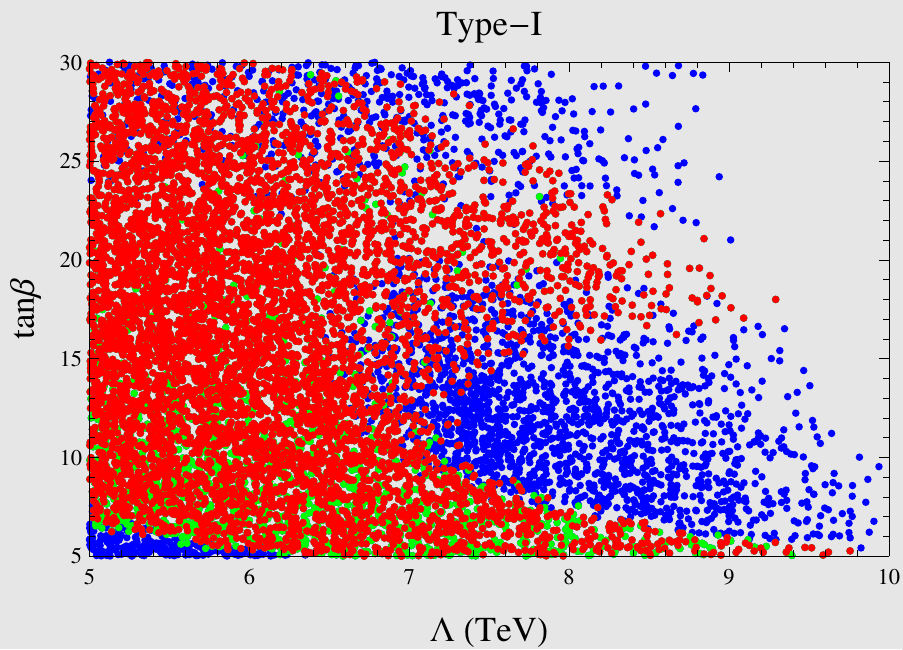}
\includegraphics[width=0.45\textwidth]{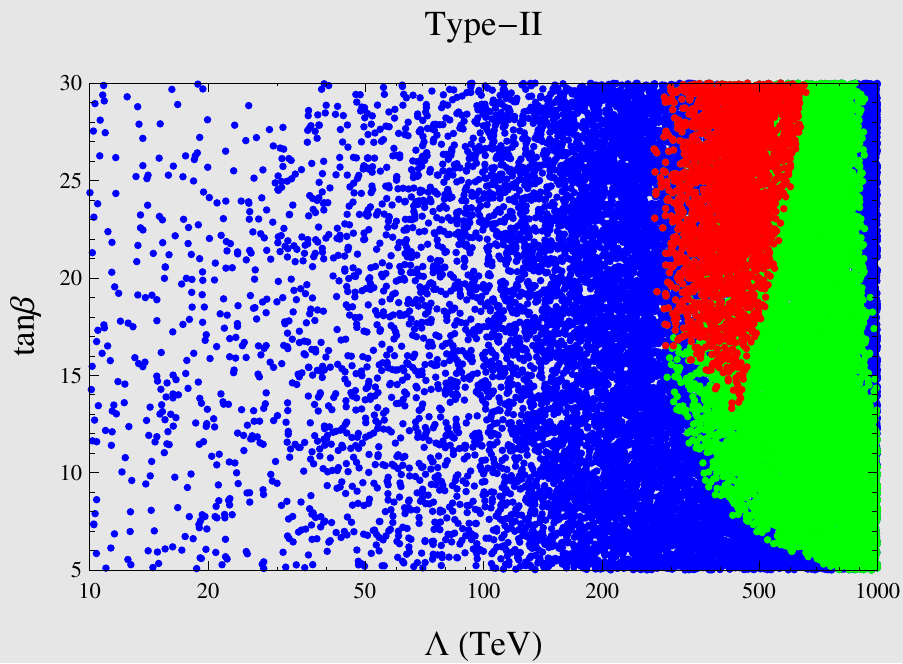}
\end{center}
\caption{Distributions of samples in the $\Lambda-\tan\beta$ plane, left (right) panel for type-I (type-II) model.}
\label{fig:sample3}
\end{figure}

\begin{figure}[!htbp]
\begin{center}
\includegraphics[width=0.45\textwidth]{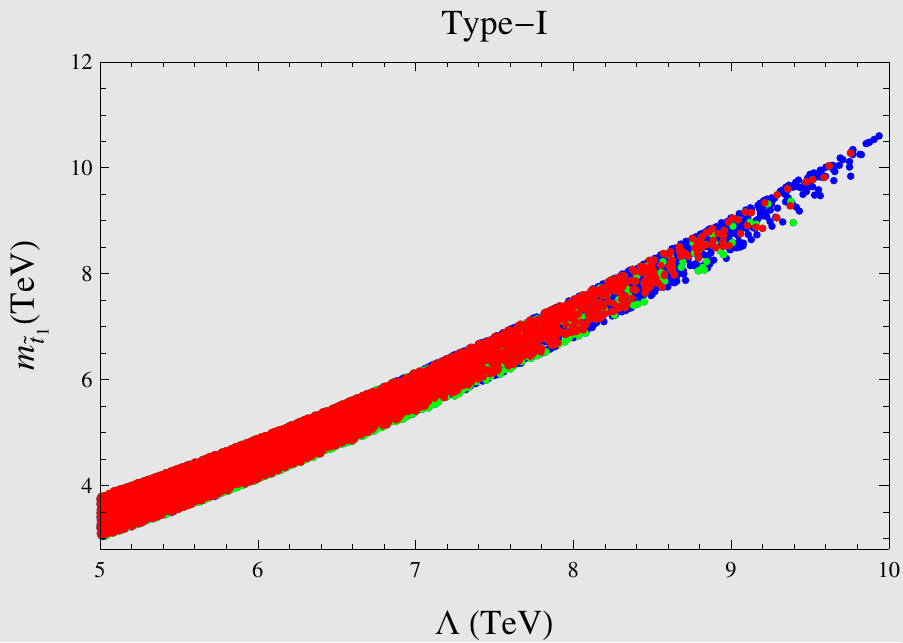}
\includegraphics[width=0.45\textwidth]{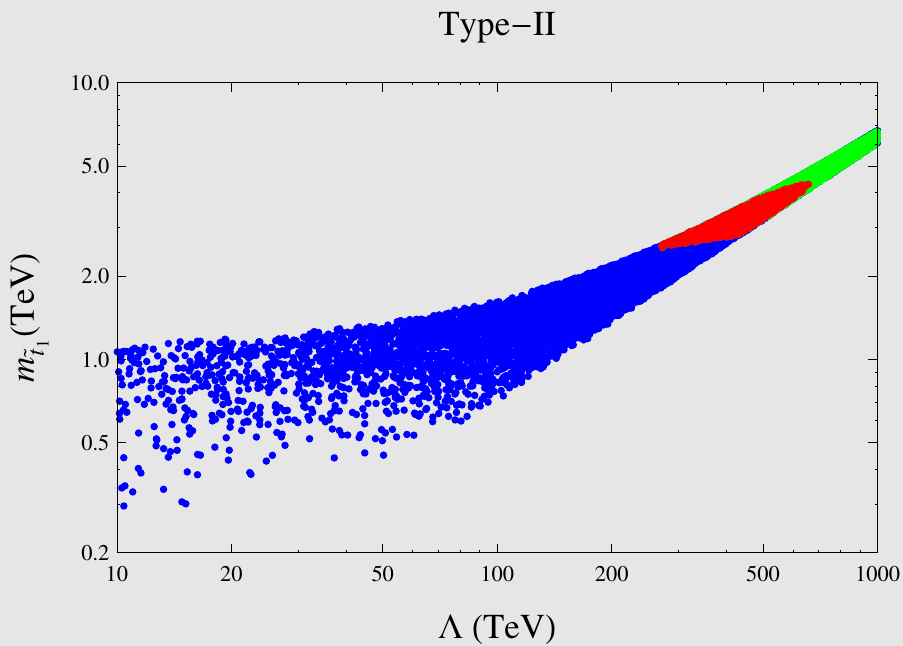}
\end{center}
\caption{Distributions of stop masses versus $\Lambda$, left (right) panel for type-I (type-II) model.}
\label{fig:stop2}
\end{figure}

\begin{figure}[!htbp]
\begin{center}
\includegraphics[width=0.45\textwidth]{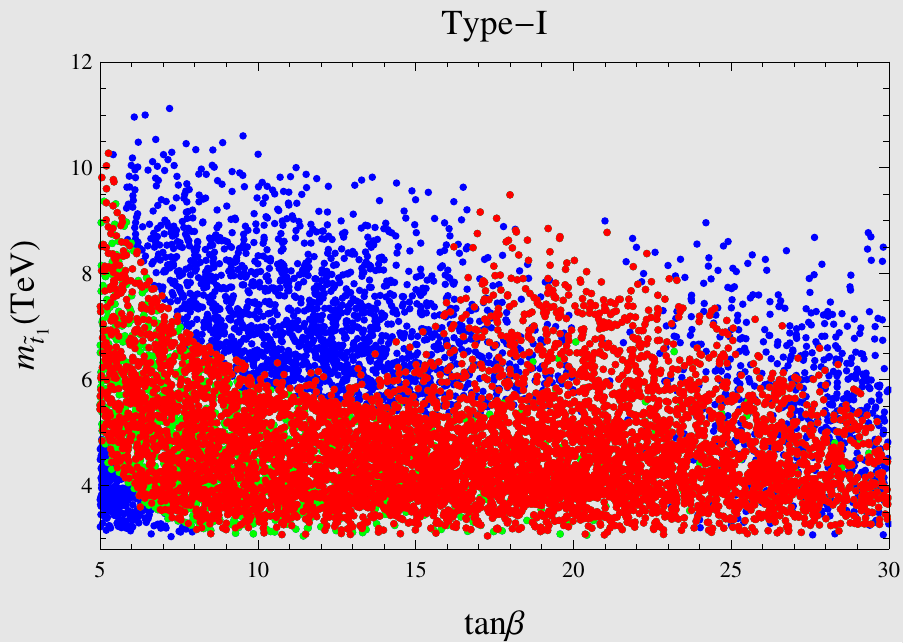}
\includegraphics[width=0.45\textwidth]{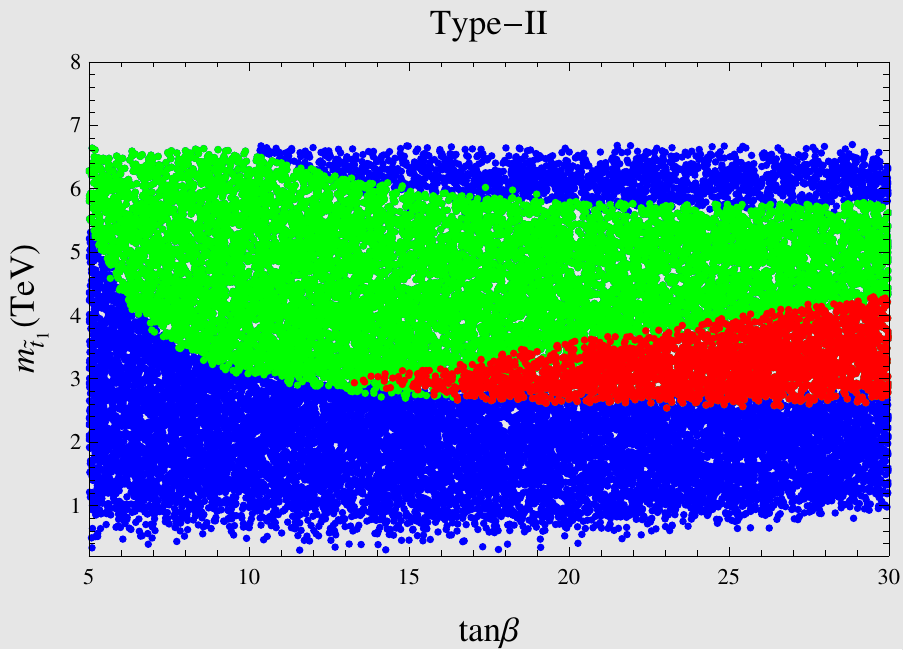}
\end{center}
\caption{Distributions of stop masses versus $\tan\beta$, left (right) panel for type-I (type-II) model.}
\label{fig:stop3}
\end{figure}

\begin{figure}[!htbp]
\begin{center}
\includegraphics[width=0.45\textwidth]{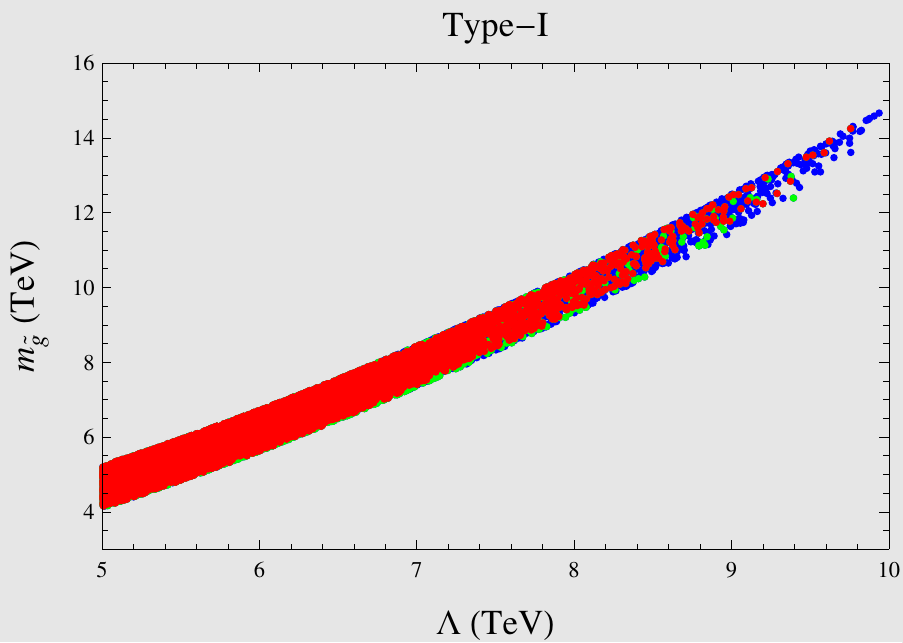}
\includegraphics[width=0.45\textwidth]{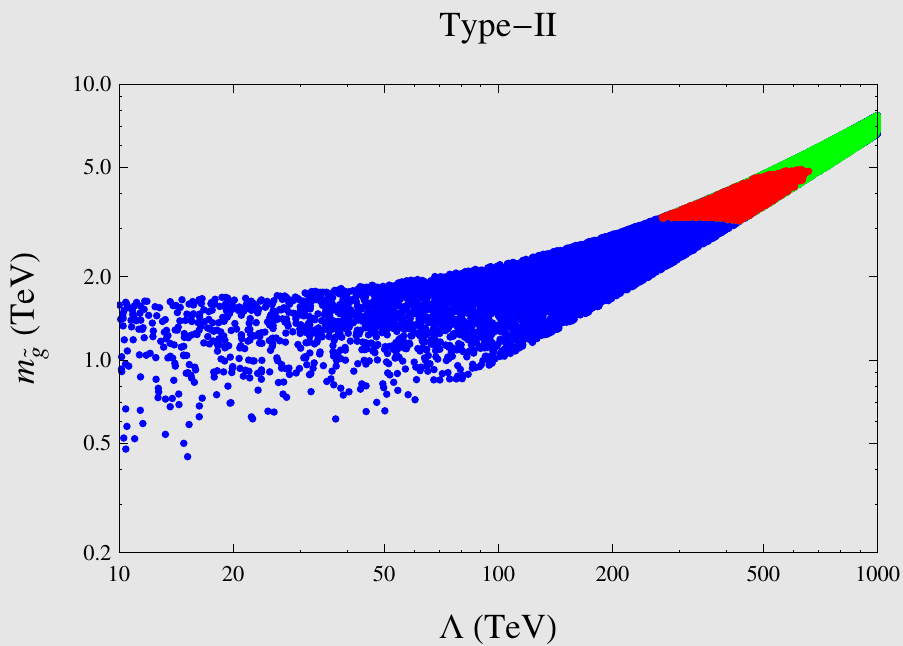}
\end{center}
\caption{Distributions of gluino masses versus $\Lambda$, left (right) panel for type-I (type-II) model.}
\label{fig:gluino2}
\end{figure}

\begin{figure}[!htbp]
\begin{center}
\includegraphics[width=0.45\textwidth]{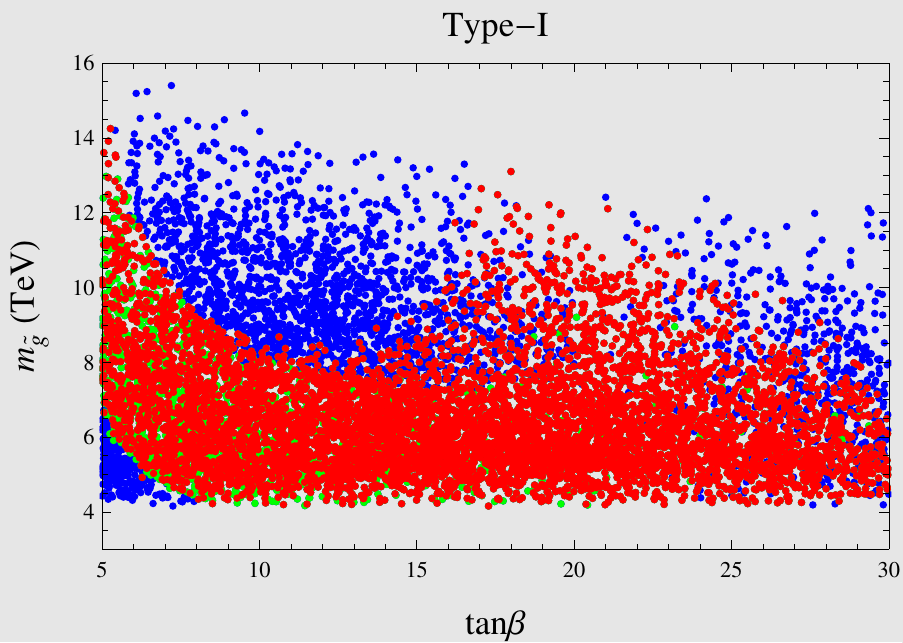}
\includegraphics[width=0.45\textwidth]{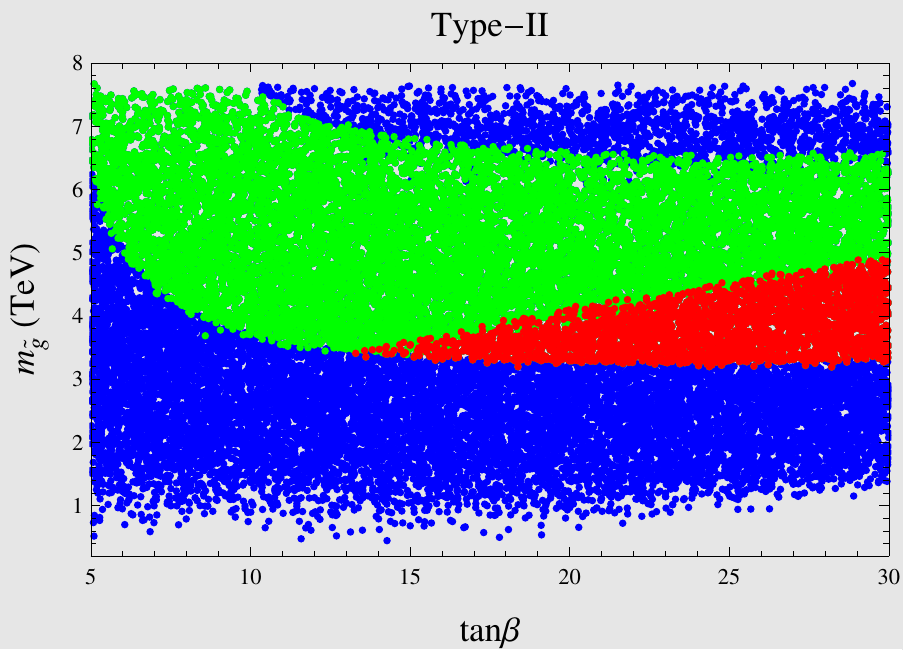}
\end{center}
\caption{Distributions of gluino masses versus $\tan\beta$, left (right) panel for type-I (type-II) model.}
\label{fig:gluino3}
\end{figure}

In order to exam the full parameter space in Eq.~(\ref{eq:parameters}), we further performed random scans over  within following regions:
\begin{equation}
\left\{\begin{array}{lll}
\Lambda\in[5\times10^3,~10^4]~{\rm GeV}\;,\quad m_0\in[10,~600]~{\rm GeV}~&\hbox{for type-I model},\nonumber\\
\Lambda\in[10^4,~10^6]~{\rm GeV}\;,\quad m_0\in[100,~1000]~{\rm GeV}~&\hbox{for type-II model},\nonumber\\
m_{1/2}\in[100,~700]~{\rm GeV}\;,\quad \tan\beta\in[5,~30]~&\hbox{for both type-I and II models}.
\end{array}\right.
\label{eq:ranges}
\end{equation}
During the scan, we applied various constraints come from collider and low-energy experiments are applied, which are given below
\begin{enumerate}
\item {\textbf{The higgs mass:} $m_h\in[123,~127]$ GeV}.
\item {\textbf{LEP bounds and B physics constraints:}
    \begin{eqnarray}
    \label{eqn:Bphysics}
    1.6\times 10^{-9} \leq{\rm BR}(B_s \rightarrow \mu^+ \mu^-)
    \leq 4.2 \times10^{-9}\;(2\sigma)~\text{\cite{CMS:2014xfa}}\;,
    \nonumber\\
    2.99 \times 10^{-4} \leq
    {\rm BR}(b \rightarrow s \gamma)
    \leq 3.87 \times 10^{-4}\;(2\sigma)~\text{\cite{Amhis:2014hma}}\;,
    \nonumber\\
    7.0\times 10^{-5} \leq {\rm BR}(B_u\rightarrow\tau \nu_{\tau})
    \leq 1.5 \times 10^{-4}\;(2\sigma)~\text{\cite{Amhis:2014hma}}\;.
    \end{eqnarray}}
    \item {\textbf{The muon $g-2$ anomaly:}
    \begin{eqnarray}
    4.7\times 10^{-10} \leq \Delta a_\mu
        \leq 52.7 \times 10^{-10}\;(3\sigma)~\text{\cite{Davier:2010nc}}\;.
    \end{eqnarray}}
    \item {\textbf{LHC constraints:}
    \begin{itemize}
    \item Gluino mass $m_{\tilde g} > 1800$ GeV~\cite{ATLAS:2016kts,CMS:2016inz},
    \item Light stop mass $m_{{\tilde t}_1} > 850$ GeV~\cite{ATLAS:2016jaa,CMS:2016inz},
    \item Light sbottom mass $m_{{\tilde b}_1}>840-1000$  GeV~\cite{Aaboud:2016nwl,CMS:2016xva},
    \item Degenerated first two generation squarks (both left-handed and right-handed) $m_{{\tilde q}}>1000-1400$ GeV~\cite{CMS:2016xva}.
    \end{itemize}}
\end{enumerate}
The corresponding results are displayed in Figs.~\ref{fig:sample1}-\ref{fig:gluino3}. In all of these figures,
blue points denote total valid samples; green points denote
samples which fulfill higgs mass, LEP bounds and B physics constraints; red points denote samples which further satisfy muon $g-2$ anomaly and LHC constraints. As one expects, the viable parameter distributions in the full parameter space basically agree with our previous discussion, {\it i.e.},
their behaviors are mainly determined by $\Lambda$ and $m_0$, while the choice of $m_{1/2}$ and $\tan\beta$ have relatively less effects. There only one feature is worth to be emphasized: comparing with  type-I model, type-II model has heavier slepton and electroweakino spectra, which then requiring a larger $\tan\beta$ to enhance their contributions in the loop in order to provide sufficient $\Delta a_{\mu}$. To be specific, for samples passing the muon $g-2$ anomaly, $\tan\beta$ cannot lower than 13.

\begin{figure}[!htbp]
\begin{center}
\includegraphics[width=0.45\textwidth]{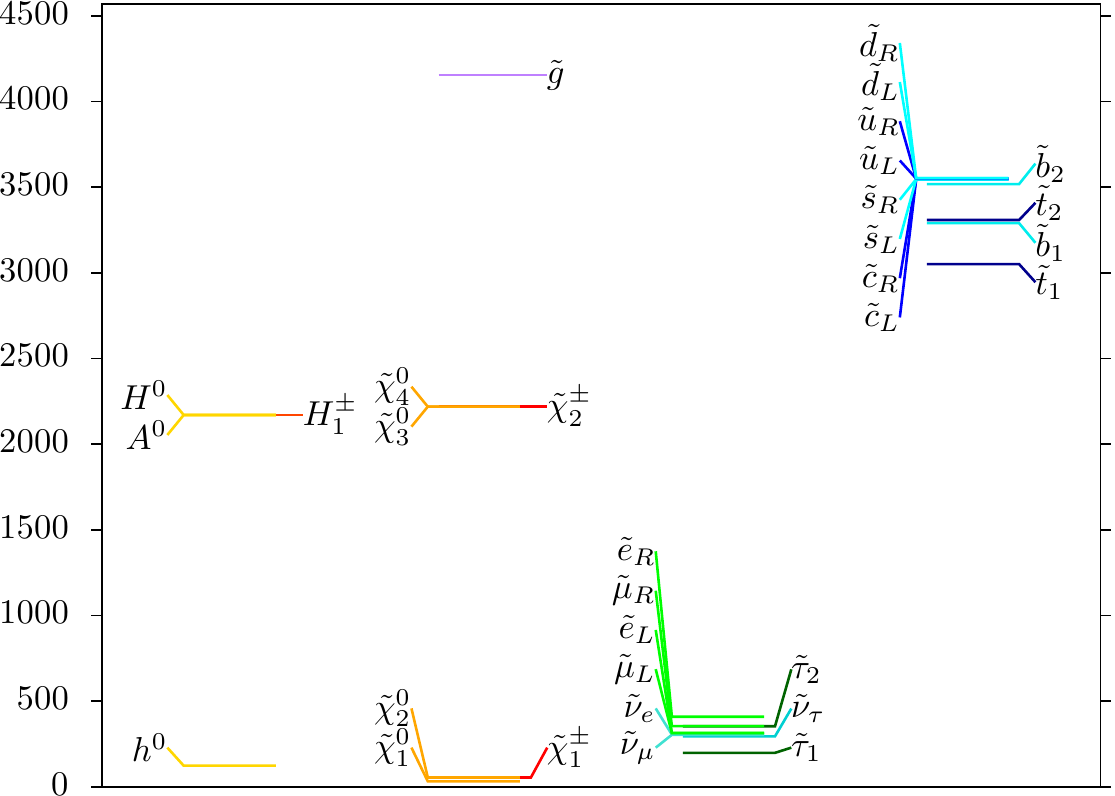}
\includegraphics[width=0.45\textwidth]{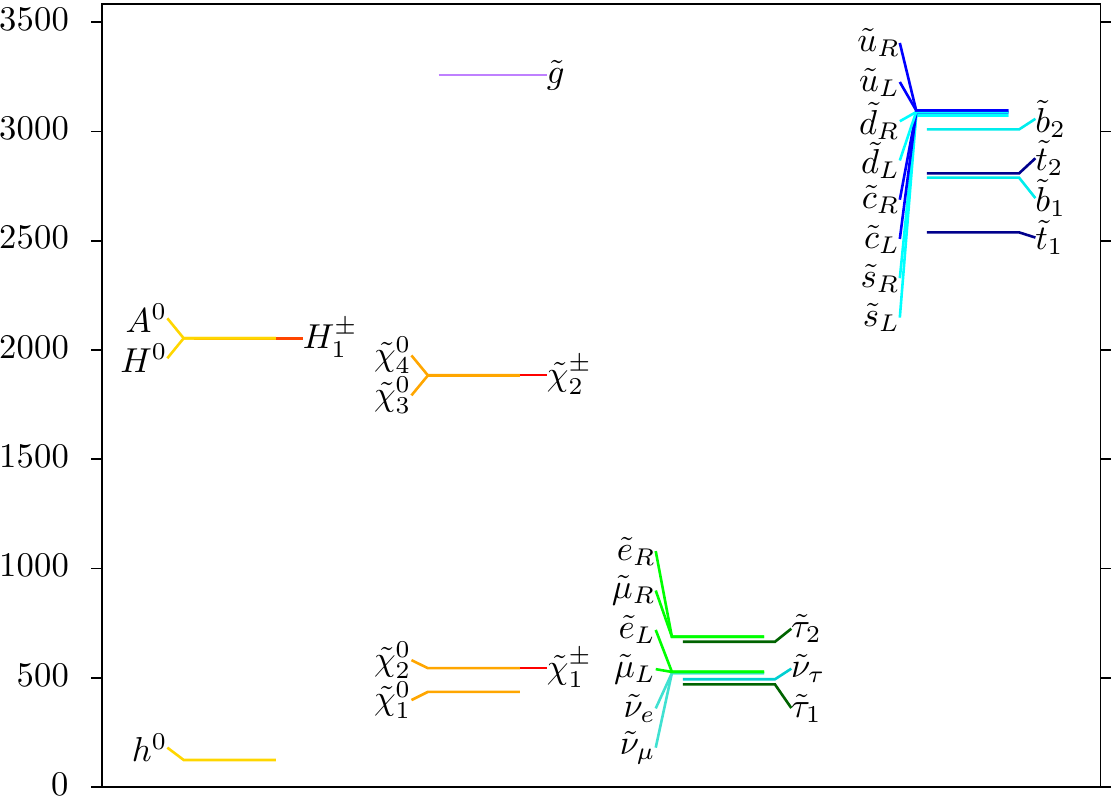}
\label{fig:bench}
\end{center}
\caption{The complete particle spectra for two benchmark points. Left (right) panel for type-I (type-II) model.}
\end{figure}

Moreover, we show benchmark points for two type models. The input parameters, important particle spectra and $\Delta a_\mu$ are shown in Table~\ref{tab:ranges}, and the corresponding complete spectra are displayed in Fig.~\ref{fig:bench}.
Finally, we would like to comment on the supersymmetry breaking soft term relations between gauge mediation
and gravity mediation for these two benchmark points.
For simplicity, we assume that gravitino mass is equal to scalar mass in gravity mediation.
In the Type-I model, the supersymmetry breaking F-term for gauge mediation is
$F_{\rm Gauge}=\Lambda M_{mess}= 5\times 10^{19}~{\rm GeV}$, while the
 supersymmetry breaking F-term for gravity mediation is
$F_{\rm Gravity}={\sqrt 3}m_0 M_{\rm Pl}= 1.45627\times 10^{21}~{\rm GeV}$.
Because there is a factor 29.1254 difference between these two F-terms,
we need some fine-tuning around 3.4\% to obtain the proper supersymmetry breaking soft terms.
In the Type-II model, the supersymmetry breaking F-term for gauge mediation is
$F_{\rm Gauge}=\Lambda M_{mess}= 2.7\times 10^{21}~{\rm GeV}$, while the
 supersymmetry breaking F-term for gravity mediation is
$F_{\rm Gravity}={\sqrt 3}m_0 M_{\rm Pl}= 4.17521\times 10^{21}~{\rm GeV}$.
So $F_{\rm Gauge}$ and $F_{\rm Gravity}$ are at the same order, and can be considered
as the same F-term. Therefore, we can explain why the
supersymmetry breaking soft terms from gauge and gravity mediations are comparable.
In other words, the Type-II model on supersymmetry breaking is indeed natural.

\begin{table*}[hbtp]
\begin{tabular}{|c|c|c|c|c|c|c|c|c|}
\hline
Model & $\Lambda$ & $m_0$ & $m_{1/2}$ & $\tan\beta$ & $m_h$ & $m_{{\tilde t}_1}$ & $m_{\tilde g}$ & $\Delta a_\mu$   \\\hline
Type-I & $5\times10^3$ & $346$ & $116$ & $17.3$ & $124$ & $3051$ & $4155$ & $1.37\times10^{-9}$ \\\hline
Type-II & $2.7\times10^5$ & $992$ & $678$ &  $23.1$  & $123$ & $2538$ & $3259$ & $6.83\times10^{-10}$  \\
\hline
\end{tabular}
\caption{The input parameters, important particle spectra and $\Delta a_\mu$ for two benchmark points. }
\label{tab:ranges}
\end{table*}

\section{Conclusion}
\label{sec:conclusion}

The muon anomalous magnetic moment has been treated as new physics guideline for a long time. If considering it as a tantalizing hint for new physics beyond the SM, we proposed a modified version of the MSSM in the framework of hybrid gauge-gravity mediation SUSY breaking, avoiding the tachyonic slepton problem in usual GMSB. We study the particle spectra with emphasize on the parameter space which can simultaneously explain the Higgs mass of $125$ GeV and the observed muon $g-2$ anomaly. We found that slepton and squarks sector can be easily split for both type-I and type-II models. In particular, type-I model is more suitable to interpret the muon anomalous magentic moment although it more relies on soft slepton masses provided by gravity mediation. In the entire of allowed parameter space, the masses of the light squarks and gluino fall into the multi-TeV mass range. It is expected to be exceeded abilities of the current and upcoming LHC SUSY direct searches and could be reached at future HE-LHC, FCC-hh and SPPC. Finally, the dark matter candidate is still the lightest neutralino and thus all of advantages in WIMP scenario can be preserved in our model.

%%%%%%%%%%%%%%%%%%%%%%%%%%%%%%%%%%%%%%%%%%%%%%%%%%%

\begin{acknowledgments}

This research was supported in part
by the Projects 11475238 and 11647601 supported
by National Natural Science Foundation of China,
and by Key Research Program of Frontier Science, CAS.

\end{acknowledgments}

\end{document}